\documentclass[aps,prb,reprint,twocolumn,superscriptaddress,floatfix,nofootinbib]{revtex4-1}
\usepackage{amsmath,amssymb,color,comment,physics}
\usepackage[makeroom]{cancel}
\usepackage[caption=false]{subfig}
\usepackage{mathrsfs}
\usepackage{graphicx}
\usepackage{subfig}
\usepackage[countmax]{subfloat}
\usepackage[normalem]{ulem}
\usepackage[english]{babel}
\usepackage{dsfont}
\usepackage[bookmarks=true,colorlinks,linkcolor=OrangeRed,urlcolor=NavyBlue,citecolor=RoyalBlue]{hyperref}
\usepackage[dvipsnames]{xcolor}

\usepackage{color}

\definecolor{mypurple}{rgb}{0.49,0.18,0.56}
\definecolor{mygold}{rgb}{0.93,0.59,0.13}
\definecolor{mygreen}{rgb}{0,0.5,0}
\definecolor{myblue}{rgb}{0,0,0.75}
\definecolor{mymagenta}{cmyk}{0,1,0,0.12}
\definecolor{mygray}{rgb}{0.5,0.5,0.5}

\usepackage{umoline}

\newcommand{\sign}[1]{\text{sgn}{#1}}

\definecolor{mypink1}{rgb}{0.858, 0.188, 0.478}

\begin{document}

\title{Dynamical phase transitions in quantum spin models with antiferromagnetic long-range interactions}
\author{Jad C.~Halimeh}
\email{jad.halimeh@physik.lmu.de}
\affiliation{INO-CNR BEC Center and Department of Physics, University of Trento, Via Sommarive 14, I-38123 Trento, Italy}
\author{Maarten Van Damme}
\affiliation{Department of Physics and Astronomy, University of Ghent, Krijgslaan 281, 9000 Gent, Belgium}
\author{Lingzhen Guo}
\affiliation{Max Planck Institute for the Science of Light, Staudtstrasse 2, 91058 Erlangen, Germany}
\author{Johannes Lang}
\email{gu53jup@pks.mpg.de}
\affiliation{Max Planck Institute for the Physics of Complex Systems, 01187 Dresden, Germany}
\author{Philipp Hauke}
\email{philipp.hauke@unitn.it}
\affiliation{INO-CNR BEC Center and Department of Physics, University of Trento, Via Sommarive 14, I-38123 Trento, Italy}

\begin{abstract}
In recent years, dynamical phase transitions and out-of-equilibrium criticality have been at the forefront of ultracold gases and condensed matter research. Whereas universality and scaling are established topics in equilibrium quantum many-body physics, out-of-equilibrium extensions of such concepts still leave much to be desired. Using exact diagonalization and the time-dependent variational principle in uniform martrix product states, we calculate the time evolution of the local order parameter and Loschmidt return rate in transverse-field Ising chains with antiferromagnetic power law-decaying interactions, and map out the corresponding rich dynamical phase diagram. \textit{Anomalous} cusps in the return rate, which are ubiquitous at small quenches within the ordered phase in the case of ferromagnetic long-range interactions, are absent within the accessible timescales of our simulations in the antiferromagnetic case, showing that long-range interactions are not a sufficient condition for their appearance. We attribute this to much weaker domain-wall binding in the antiferromagnetic case. For quenches across the quantum critical point, \textit{regular} cusps appear in the return rate and connect to the local order parameter changing sign, indicating the concurrence of two major concepts of dynamical phase transitions. Our results consolidate conclusions of previous works that a necessary condition for the appearance of anomalous cusps in the return rate after quenches within the ordered phase is for topologically trivial local spin flips to be the energetically dominant excitations in the spectrum of the quench Hamiltonian. Our findings are readily accessible in modern trapped-ion setups, and we outline the associated experimental considerations.
\end{abstract}
\date{\today}
\maketitle
\tableofcontents
\section{Introduction}
The understanding of equilibrium classical and quantum phase transitions is rather well-established.\cite{Cardy_book,Sachdev_book} Not only are theoretical tools such as renormalization group \cite{Wilson1971a} able to accurately predict equilibrium critical points and exponents, but also modern ultracold-atom experiments can realize quantum phase transitions with great reliability.\cite{Greiner2002,Lewenstein_review,Bloch2008,Simon2011} A foundational breakthrough in this field has been the discovery of equilibrium universality classes, which characterize a set of different models that share the same critical exponents at a continuous phase transition.\cite{Cardy_book,Sachdev_book} In other words, these models, though perhaps quite different away from the phase transition, behave very similarly in its vicinity.

The situation is not as well developed in out-of-equilibrium physics, even though impressive experimental advances in quantum simulators have allowed unprecedented control in the observation of various intriguing out-of-equilibrium phenomena such as prethermalization,\cite{Gring2012,Langen2015,Neyenhuis2017} gauge-theory dynamics,\cite{Martinez2016,Bernien2017,Dai2017,Klco2018,Kokail2019,Schweizer2019,Goerg2019,Mil2020,Yang2020} time crystals,\cite{Choi2017,Zhang2017timecrystal,Rovny2018,Smits2018} Kibble-Zurek mechanism,\cite{Xu2014,Anquez2016,Clark2016,Cui2016,Keesling2019} dynamical phase transitions,\cite{Jurcevic2017,Zhang2017dpt,Flaeschner2018,Muniz2020} many-body localization,\cite{Choi2016,Smith2016} and many-body dephasing.\cite{Kaplan2020} Even though dynamical universality classes have been discovered for classical systems, \cite{Hohenberg1977} quantum many-body models still lack a clear classification of dynamical universality. 

Recently, a connection has been made in out-of-equilibrium quantum many-body systems to equilibrium criticality. Specfically, the concept of \textit{dynamical quantum phase transitions}\cite{Heyl2013,Heyl2014,Heyl2015,Heyl_review,Zvyagin2016,Mori2018} (DQPT; also known as DPT-II\cite{footnote1}) has been introduced to offer a dynamical analog of the thermal free energy in the form of the Loschmidt return rate
\begin{align}\label{eq:ReturnRate}
r(t)=-\lim_{N\to\infty}\frac{1}{N}\ln\big\lvert \bra{\psi_0}e^{-iHt}\ket{\psi_0}\big\rvert^2,
\end{align}
where complexified time stands for inverse temperature. Here, $N$ is the size of the system under study, $H$ its Hamiltonian, and $\ket{\psi_0}$ the initial state. The time $t_\mathrm{c}$ at which $r(t)$ exhibits a nonanalyticity is a \textit{critical time} at which a \textit{dynamical quantum phase transition} occurs, much the same way as in equilibrium where a  thermal phase transition occurs at the critical temperature at which the thermal free energy has a nonanalyticity. The theory of thermal phase transitions is thus extended into the far-from-equilibrium realm, wherein DPT-II serves as a formal concept of dynamical phase transitions. Further extensions of universal scaling behavior in the vicinity of cusps in the return rate have also been recently explored.\cite{Halimeh2019b,Wu2019,Wu2020a,Wu2020b,Trapin2020,Halimeh2020} A rigorous treatment of DPT-II can be found in several recent reviews,\cite{Zvyagin2016,Heyl_review,Mori2018} along with experimental observations of them.\cite{Jurcevic2017,Zhang2017dpt,Flaeschner2018}

The Loschmidt return rate~\eqref{eq:ReturnRate} can exhibit nonanalytic \textit{cusps} under certain conditions.\cite{Silva2008} In the seminal work of Ref.~\onlinecite{Heyl2013}, it has been shown that quenches in the nearest-neighbor transverse-field Ising chain (NN-TFIC) give rise to nonanalyticities in the Loschmidt return rate only when carried out across the equilibrium quantum critical point $h_\mathrm{c}^\mathrm{e}$. Early assumptions that the appearance of such cusps in the return rate may be specific to integrable models mappable onto two-band free fermionic systems have been refuted through numerical studies\cite{Karrasch2013} in the time-dependent density matrix renormalization group \cite{White1992,White1993,Uli2005,Schmitteckert2004,Feiguin2005,GarciaRipoll2006,McCulloch2007} based on matrix product states.\cite{Oestlund1995,Uli2010} Indeed, the next-nearest-neighbor transverse-field and the nearest-neighbor tilted-field Ising chains, both nonintegrable models, exhibit nonanalytic Loschmidt return rates in the wake of certain quenches.\cite{Karrasch2013} Additionally, quenching across the equilibrium quantum critical point has been found to be neither a necessary nor sufficient condition for cusps to appear in the Loschmidt return rate.\cite{Vajna2014,Andraschko2014,Jafari2019} Even in two-band free fermionic models with long-range couplings, \cite{Dutta2017,Defenu2019,Uhrich2020} a quench is found to cause cusps in the Loschmidt return rate only when it crosses a dynamical critical point $h_\mathrm{c}^\mathrm{d}\leq h_\mathrm{c}^\mathrm{e}$, but it is not necessary for the quench to go across $h_\mathrm{c}^\mathrm{e}$.

Naturally, the question then arises of how DPT-II is connected to the Landau-type dynamical phase transition (DPT-I) characterized by the dynamics of a local order parameter in the wake of a quench.\cite{Sciolla2010,Sciolla2011,Sciolla2013} Considering the case of an ordered initial state, the local order parameter can either asymptotically decay to a constant value without changing sign for small quenches to $h< h_\mathrm{c}^\mathrm{d}$, or it oscillates around zero with an amplitude that vanishes in the long-time limit for large quenches to $h> h_\mathrm{c}^\mathrm{d}$. \cite{Altman2002,Calabrese2006,Calabrese2007a,Barmettler2009,Halimeh2017b,Calabrese2011,Calabrese2012} In case the model has a finite-temperature phase transition, the former (latter) case coincides with an ordered (disordered) long-time steady state. Numerical studies have indicated that $h_\mathrm{c}^\mathrm{d}$ is the same for DPT-I and DPT-II particularly in the case of ferromagnetic (FM) long-range interactions, \cite{Halimeh2017,Zauner2017,Homrighausen2017,Lang2017,Lang2018,Zunkovic2018,Piccitto2019} although this depends on the specific definition of the phases of DPT-II.\cite{footnote2} This picture is also found to persist even at finite temperature. For example, quenches in the NN-TFIC at finite temperature will always give rise to an analytic return rate,\cite{Abeling2016} which coincides with the absence of an equilibrium phase transition in this model at finite temperature. On the other hand, the fully connected transverse-field Ising model, also known as the Lipkin-Meshkov-Glick (LMG) model, which supports a finite-temperature phase transition in equilibrium, exhibits two different kinds of cusps depending on whether one quenches below or above a dynamical critical point $h_\text{c}^\text{d}(h_\text{i},T)$ from the ordered phase, where this dynamical critical point depends on the initial value of the transverse-field strength $h_\text{i}$ and the temperature $T$ at which the initial state is prepared.\cite{Homrighausen2017,Lang2017,Lang2018,Sehrawat2020} This is also found to coincide with a long-time ferromagnetic or paramagnetic steady state, respectively, and is directly related to the LMG model hosting an equilibrium phase transition at a temperature $T<T_\mathrm{c}$, where $T_\mathrm{c}$ is its thermal critical point. Cusps arising in the return rate of the LMG model for quenches below $h_\mathrm{c}^\mathrm{d}(h_\text{i},T)$ within the ordered phase, where $T$ is the temperature at which the initial state is prepared, are called \textit{anomalous} and they do not correspond to any zeros in the dynamics of the order parameter. Their \textit{regular} counterparts occur when the quench goes from the ordered phase to above $h_\mathrm{c}^\mathrm{d}(h_\text{i},T)$, and just as in the traditional case of the NN-TFIC at $T=0$, these cusps correspond to zeros in the dynamics of the order parameter. Anomalous cusps have been shown to occur when the quench Hamiltonian's spectrum in the ordered phase energetically favors local spin excitations, such as in the case of the LMG model,\cite{Homrighausen2017,Lang2017,Lang2018} two-dimensional quantum Ising models, \cite{Hashizume2018,Hashizume2020} and transverse-field Ising chains with sufficiently long-range FM interactions.\cite{Halimeh2017,Zauner2017,Halimeh2018a}

In this work, we extend the connection between DPT-I and DPT-II to transverse-field Ising chains with \textit{antiferromagnetic} (AF) power-law interactions. Given their experimental feasibility and more natural implementation in trapped ions relative to their FM counterparts,\cite{Britton2012,Islam2013,Richerme2014,Jurcevic2014,Jurcevic2015,Smith2016,Jurcevic2017,Zhang2017dpt} it is important to understand the dynamical critical properties of these systems to guide and complement modern ion-trap setups in their investigations. Moreover, when interactions are long-range, frustration arises in the AF case, but not in the FM case. It is therefore an interesting question in its own right what the effect of frustration is on quench dynamics in the AF-TFIC. As we demonstrate using extensive numerical calculations based on tensor network methods and exact diagonalization, long-range interactions are not a sufficient condition for anomalous cusps to appear, such that frustration considerably modifies the phenomenology of DQPTs.

The rest of the paper is organized as follows: In Sec.~\ref{sec:model}, we discuss the paradigmatic quantum Ising chain with power-law interactions and its equilibrium features both in the AF and FM cases, including analytic and numerical investigations of the underlying domain-wall interactions. Our numerical results obtained from infinite matrix product states (iMPS) and exact diagonalization (ED) are then presented in Sec.~\ref{sec:numerics}. Experimental considerations of our findings are discussed in Sec.~\ref{sec:experimental}. We provide concluding remarks and an outlook in Sec.~\ref{sec:conclusion}. In Appendix~\ref{sec:kink} we discuss the two-kink model. Further details on our iMPS framework are provided in Appendix~\ref{sec:NumSpec}. A detailed explanation of our ED implementation can be found in Appendix~\ref{sec:Neel}.

\begin{figure}[htp]
	\centering
	\includegraphics[width=.48\textwidth]{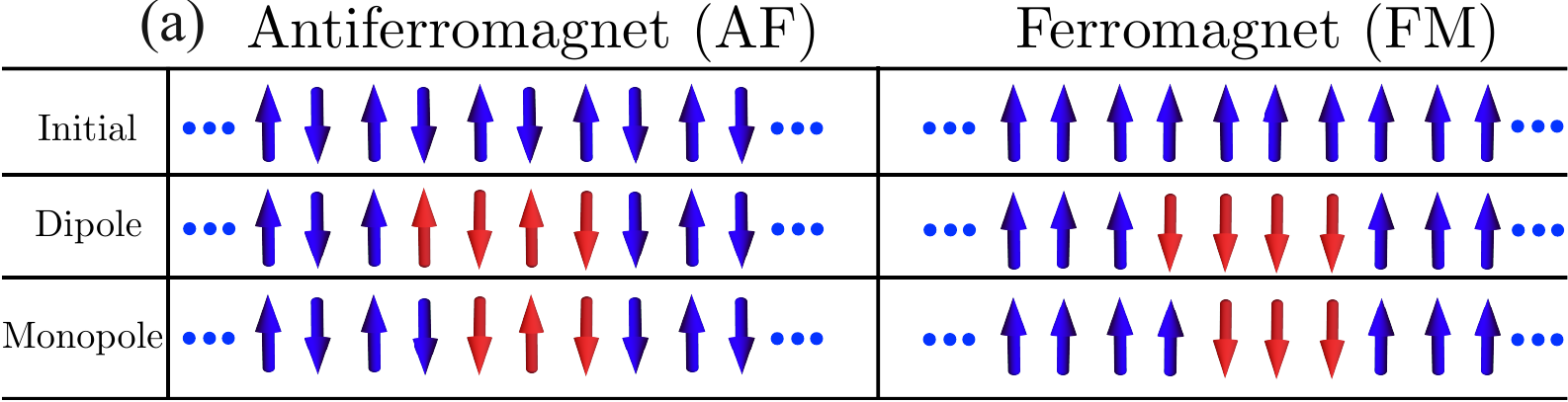}\\
	\vspace{1.5mm}
	\includegraphics[width=.48\textwidth]{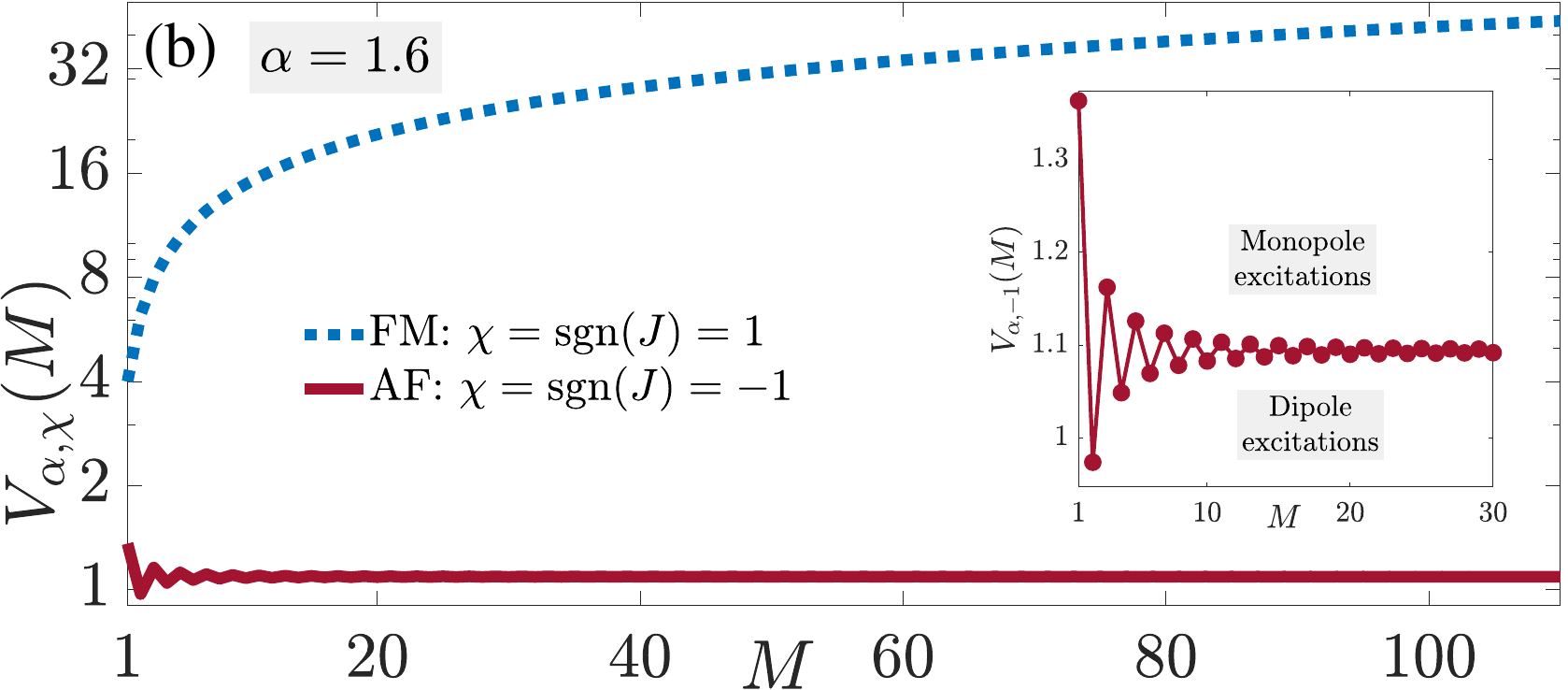}\\
	\includegraphics[width=.48\textwidth]{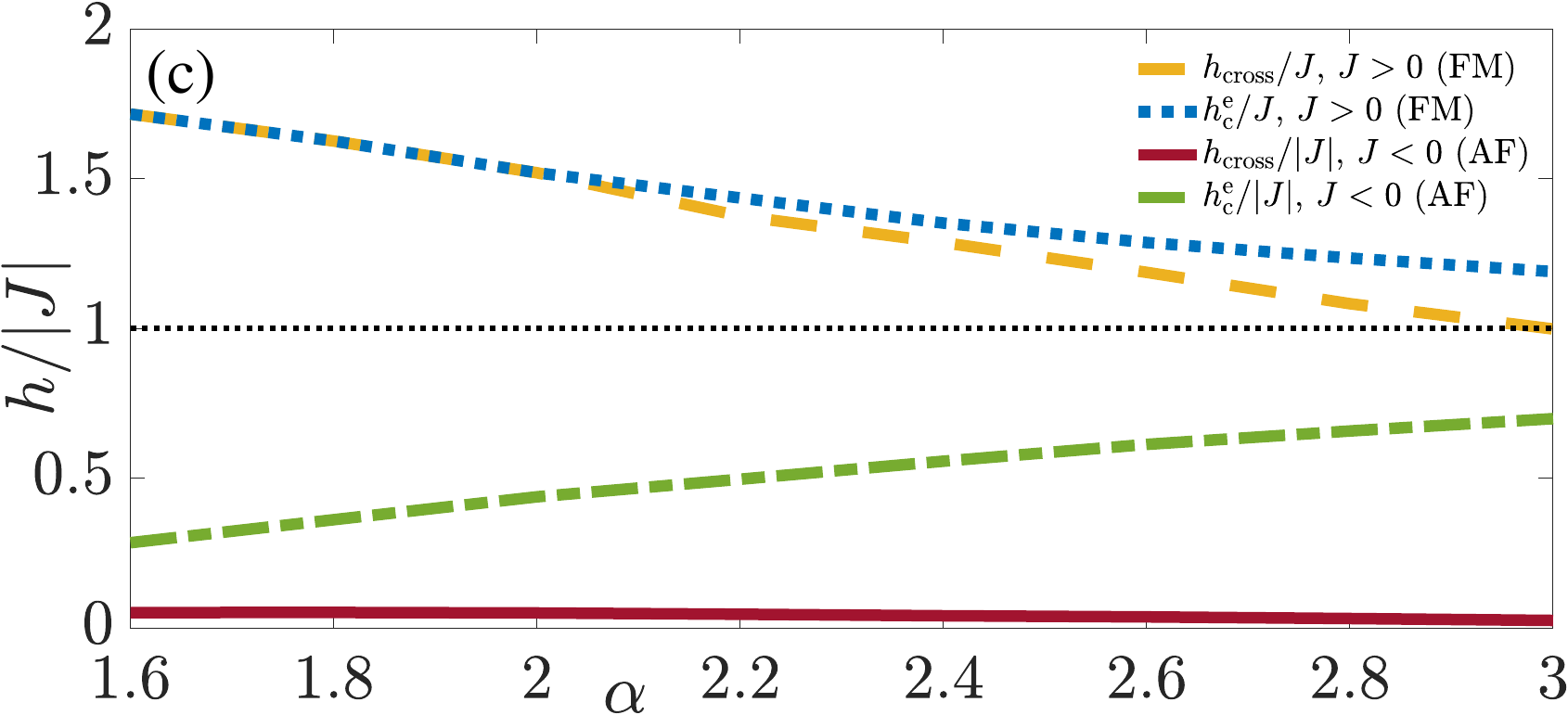}
	\caption{(Color online). (a) Initial states and single-domain excitations with a domain of length $M$ spins. We call such excitations monopoles (dipoles) if $M$ is odd (even). We note that even though we initially prepare the system in a ground state of Eq.~\eqref{eq:H} at field strength $h_\text{i}=0$, we have checked that our conclusions are independent of the choice of $h_\text{i}<h_\text{c}^\text{e}$. (b)  The domain potential $V_{\alpha,\chi}(M)$, see Eq.~\eqref{eq:potential}, denotes the cost of a domain of $M$ sites with respect to a ground state of Eq.~\eqref{eq:H} at zero transverse-field strength. Shown here for $\alpha=1.6$, this potential is significantly greater in the case of ferromagnetic (FM, $J>0$) compared to antiferromagnetic (AF, $J<0$) interactions. Distinctly from the FM case, monopole excitations under AF interactions are energetically more costly due to frustration, a feature absent in the case of FM interactions. (c) Equilibrium phase diagram of the power-law interacting ($\propto -J/r^\alpha$ with $r$ inter-spin distance) transverse-field Ising chain (TFIC), given in Eq.~\eqref{eq:H} with FM or AF interactions. The $\alpha$-dependent equilibrium quantum critical point $h_\text{c}^\text{e}(\alpha)$ separates an ordered phase at $h<h_\text{c}^\text{e}$ from a paramagnetic phase at $h>h_\text{c}^\text{e}$. The \textit{crossover} field strength $h_\text{cross}(\alpha)$, also $\alpha$-dependent, is an upper bound with regards to domain-wall binding. At sufficiently small $h<h_\text{cross}$, it is always energetically favorable in the case of FM interactions for domain walls to bind. When interactions are AF this applies only in case of dipole single-domain states, whereas monopole domains are energetically favored to either grow into larger monopole domains or decay into dipoles. Domain-wall binding is particularly prominent in the TFIC with FM power-law interactions where for small $\alpha$ (i.e., long-range interactions) we find that $h_\text{cross}\approx h_\text{c}^\text{e}$. On the other hand, in the TFIC with AF power-law interactions, domain-wall binding is very weak with a much smaller value of $h_\text{cross}$, where we also find that $h_\text{cross}\ll h_\text{c}^\text{e}$ for the range of $\alpha$ accessible in our matrix product state calculations.}
	\label{fig:statics} 
\end{figure}

\section{Model and its equilibrium physics}\label{sec:model}
The power-law interacting TFIC is given by the Hamiltonian
\begin{align}\label{eq:H}
H=-\frac{J}{\mathcal{K}_\alpha}\sum_{j<\ell}\frac{\sigma_j^z\sigma_\ell^z}{\lvert\ell-j\rvert^\alpha}-h\sum_j\sigma_j^x,
\end{align}
where $\sigma_j^{\{x,y,z\}}$ are the Pauli matrices on site $j$, $h$ is the transverse-field strength, $\alpha$ is the interaction exponent, and $J$ is the spin coupling constant. We have also included the Kac normalization\cite{Kac1963}
\begin{align}
\mathcal{K}_\alpha=\lim_{N\to\infty}\frac{1}{N-1}\sum_{m=1}^N\frac{N-m}{m^\alpha},
\end{align}
which ensures energy extensivity for $0\leq\alpha\leq 1$. AF interactions are realized for $J<0$, while interactions are FM when $J>0$. Generically, this model is nonintegrable for finite $\alpha>0$, with two integrable points at $\alpha=0$ giving rise to the LMG model in the limit of infinite-range interactions, and $\alpha\to\infty$ leading to the NN-TFIC.\cite{Russomanno2020,Defenu2020}

The excitations of the TFIC are either topological in the form of domain-wall states, or topologically trivial in the case of local spin flips, and in the FM case it has been shown that there is a crossover between these excitations at sufficiently long-range interactions based on the quasiparticle Ansatz.\cite{Vanderstraeten2018} In Refs.~\onlinecite{Halimeh2018a,Liu2019}, long-range interactions have been shown to strongly influence quench dynamics and dynamical criticality when binding of domain walls becomes energetically favorable in the spectrum of the quench Hamiltonian. Indeed, in the presence of FM power-law interactions at small transverse-field strength, single-domain states are energetically favored to have small domains, see Fig.~\ref{fig:statics}. This domain-wall binding leads to constrained dynamics with very slow decay of the  order parameter in the wake of a quench, even when the model is in the short-range equilibrium universality class ($\alpha\geq3$).\cite{Halimeh2017b,Liu2019} Domain-wall coupling has also been shown to be a necessary condition for the appearance of anomalous cusps in the Loschmidt return rate.\cite{Halimeh2017,Defenu2019} This intriguing phenomenon also exists in quantum Ising chains with exponentially decaying interactions.\cite{Halimeh2018a} When bound domain walls dominate the low-energy spectrum of the quench Hamiltonian, then even when the model is of the short-range universality class in equilibrium, small quenches from the ordered phase do not lead to an exponential decay in the order parameter, as in the case of nearest-neighbor interactions where domain walls propagate freely.\cite{Calabrese2011,Calabrese2012} Instead, the order parameter shows persistent long-time order and oscillations, as has been shown in recent exact diagonalization\cite{Liu2019} and iMPS studies.\cite{Halimeh2017b,Halimeh2018a} Only at sufficiently large $h$ does the effect of domain-wall binding vanish, as then domain walls are not attractive just as in the case of the NN-TFIC, where domain walls freely propagate at any value of $h$ below the equilibrium critical point.  

This domain-wall binding effect is also present in the AF-TFIC, but there is additionally domain-wall repulsion, depending on the type of single-domain state considered. As depicted in Fig.~\ref{fig:statics}(a), we can define in the cases of both the FM-TFIC and the AF-TFIC two kinds of single-domain states: \textit{dipole} excitations, where the domain in between the two domain walls is of an even length $M$, and \textit{monopole} excitations when $M$ is odd. In order to better understand how these two types of excitations behave under FM or AF interactions, it is useful to consider the low-energy regime of the system in Eq.~\eqref{eq:H} by mapping it onto the two-kink model\cite{Coldea2010,Rutkevich2010,Liu2019} (see Appendix~\ref{sec:kink}). Inherently, the latter restricts the Hilbert space to just single-domain states, which is a valid approximation when $\lvert h\rvert\ll\lvert J\rvert/\mathcal{K}_\alpha$ in Eq.~\eqref{eq:H}. The energetic cost of a single-domain state with respect to the corresponding ground state of Eq.~\eqref{eq:H} at zero transverse-field strength is given by the \textit{domain potential}
\begin{align}\label{eq:potential}
	V_{\alpha,\chi}(M)=\frac{4J}{\mathcal{K}_\alpha}\sum_{\ell=1}^M\sum_{r=\ell}^{N\to\infty}\frac{\chi^r}{r^\alpha},
\end{align}
where $M$ is the length of the domain and $\chi=\sign(J)$. 

The energetic cost of two domain walls separated by $M$ sites is strongly influenced by whether the interactions are FM ($\chi=1$) or AF ($\chi=-1$), particularly when they are long-range. The staggering due to $\chi=-1$ in the AF case leads to a much smaller $V_{\alpha,\chi}(M)$ than in the FM case for the same $\alpha$ and $M$: $V_{\alpha,-1}(M)\ll V_{\alpha,+1}(M)$. Indeed, this is demonstrated in Fig.~\ref{fig:statics}(b) for $\alpha=1.6$. It will be interesting to see how the much weaker domain-wall binding in the case of AF interactions will influence the quench dynamics of the AF-TFIC compared to its FM counterpart. Another major difference between the domain potentials of the FM and AF cases is that in the former dipole and monopole excitations behave the same, insomuch that they favor smaller domains at zero transverse-field strength. In the AF case, this holds true for dipole domains, whereas monople domains are energetically favored to be large; see inset of Fig.~\ref{fig:statics}(b). In the regime of small quenches, dipole excitations are expected to play a more dominant role in the case of AF interactions, as monopole excitations are high-energy states.

We summarize the equilibrium physics of the AF-TFIC along with that of its FM counterpart for comparison. The only difference from its FM counterpart is the sign of $J$ in Eq.~\eqref{eq:H}. As we have seen, this seemingly trivial difference leads to vastly different equilibrium properties and phase diagrams,\cite{Koffel2012,Jaschke2017} as illustrated in Fig.~\ref{fig:statics}(c). Whereas the FM-TFIC hosts a finite-temperature phase transition for $\alpha<2$,\cite{Landau2013,Dyson1969,Thouless1969,Dutta2001} the AF-TFIC does not, regardless of the value of $\alpha$. Long-range interactions are relevant for $\alpha<3$ in the FM case,\cite{Dutta2001} while in the AF-TFIC they are relevant for $\alpha\leq2.25$.\cite{Koffel2012} Similarly to its FM counterpart, at zero temperature the AF-TFIC has an $\alpha$-dependent equilibrium quantum critical point $h_\text{c}^\text{e}(\alpha)$, which however increases from zero at $\alpha=0$ to $|J|$ in the nearest-neighbor limit $\alpha\to\infty$, whereas in the FM case it is $2J$ at $\alpha=0$ and decreases monotonically to $J$ in the nearest-neighbor limit.\cite{Vidal2004,Chen2009,Jaschke2017} The difference between both models is indeed striking at small $\alpha$, as shown by the MPS calculations in Fig.~\ref{fig:statics}(c) of the equilibrium quantum critical point $h_\mathrm{c}^\mathrm{e}$ and the \textit{crossover} field strength $h_\mathrm{cross}$. We define the latter as the value of the transverse-field strength below which it is energetically more favorable for two domain walls to be \textit{bound}, rather than infinitely separated.\cite{Halimeh2018a} In the AF case, this applies only for dipole excitations, because their monopole counterparts are repulsive. In this sense, $h_\mathrm{cross}$ serves as an upper bound for domain-wall binding in the FM-TFIC and for dipole excitations in the AF-TFIC. 

The equilibrium quantum critical points of the FM-TFIC and AF-TFIC separate further the smaller $\alpha$ is, whereas in the short-range limit $\alpha\geq3$ they approach $|J|$, which is the equilibrium quantum critical point of the NN-TFIC irrespective of whether interactions are FM or AF. Common to both models, $h_\mathrm{cross}$ approaches $h_\mathrm{c}^\mathrm{e}$ at longer interaction ranges. Within the accessible $\alpha$ values in our MPS calculations, we find that $h_\mathrm{cross}\approx h_\mathrm{c}^\mathrm{e}$ for $\alpha\lesssim2$ in case of FM interactions, whereas for the AF-TFIC $h_\mathrm{cross}\ll h_\mathrm{c}^\mathrm{e}$ down to the smallest finite $\alpha$ we achieve. As we will see in Sec.~\ref{sec:nonintegrable}, this will lead to dynamical criticality  in the AF-TFIC that is fundamentally different from that of the FM-TFIC.

\section{Quench dynamics}\label{sec:numerics}
We now present our numerical results for the time evolution of the local order parameter
\begin{align}\label{eq:OP}
	\mathcal{M}(t)=\lim_{N\to\infty}\frac{1}{N}\sum_{j=1}^N(-1)^j\bra{\psi_0}e^{iHt}\sigma_j^ze^{-iHt}\ket{\psi_0},
\end{align} 
and return rate~\eqref{eq:ReturnRate} in the AF-TFIC starting in a ground state of Eq.~\eqref{eq:H} at an initial transverse-field strength $h_\mathrm{i}=0$, and then quenching to a final value $h>0$. For $\alpha>0$, the ground-state manifold of the AF-TFIC for $h_\text{i}<h_\mathrm{c}^\mathrm{e}$ is doubly degenerate, consisting of the two N\'eel states $\ket{\cdots\uparrow_j\downarrow_{j+1}\cdots}$ and $\ket{\cdots\downarrow_j\uparrow_{j+1}\cdots}$ at $h_\mathrm{i}=0$. In the fully connected limit of $\alpha=0$, the ground-state manifold is infinitely degenerate, growing with system size $N$ as $\mathcal{O}(N!/[(N/2)!]^2)$. Even though in this work we present results only for $h_\text{i}=0$, we have checked that our conclusions are valid also for other values of $h_\text{i}<h_\text{c}^\text{e}$. We do not present results for $h_\text{i}>h_\text{c}^\text{e}$, as the picture is qualitatively identical to that of the FM-TFIC already presented in Ref.~\onlinecite{Halimeh2017}.

\subsection{Nonintegrable model at finite $\alpha>1$}\label{sec:nonintegrable}
 Using iMPS in the framework of the time-dependent variational principle (TDVP),\cite{Haegeman2011,Haegeman2016,Vanderstraeten2019} we calculate the time evolution of the local order parameter~\eqref{eq:OP} and the Loschmidt-echo return rate~\eqref{eq:ReturnRate}. We restrict our calculations to $\alpha\geq1.6$ due to numerical overhead; see Appendix~\ref{sec:LR} for details. We use the trick of backward time evolution in order to double the longest evolution time accessible for the return rate, although this is not possible for the local order parameter.\cite{Halimeh2017} This is why in the following iMPS results $r(t)$ achieves double the maximal evolution time reached in $\mathcal{M}(t)$; see Appendix~\ref{sec:backward} for details. Our most demanding calculations achieve convergence at an evolution time-step $\tau=0.005/J$ with a maximal bond dimension $\mathcal{D}=300$.

\begin{figure}[htp]
	\centering
	\includegraphics[width=.48\textwidth]{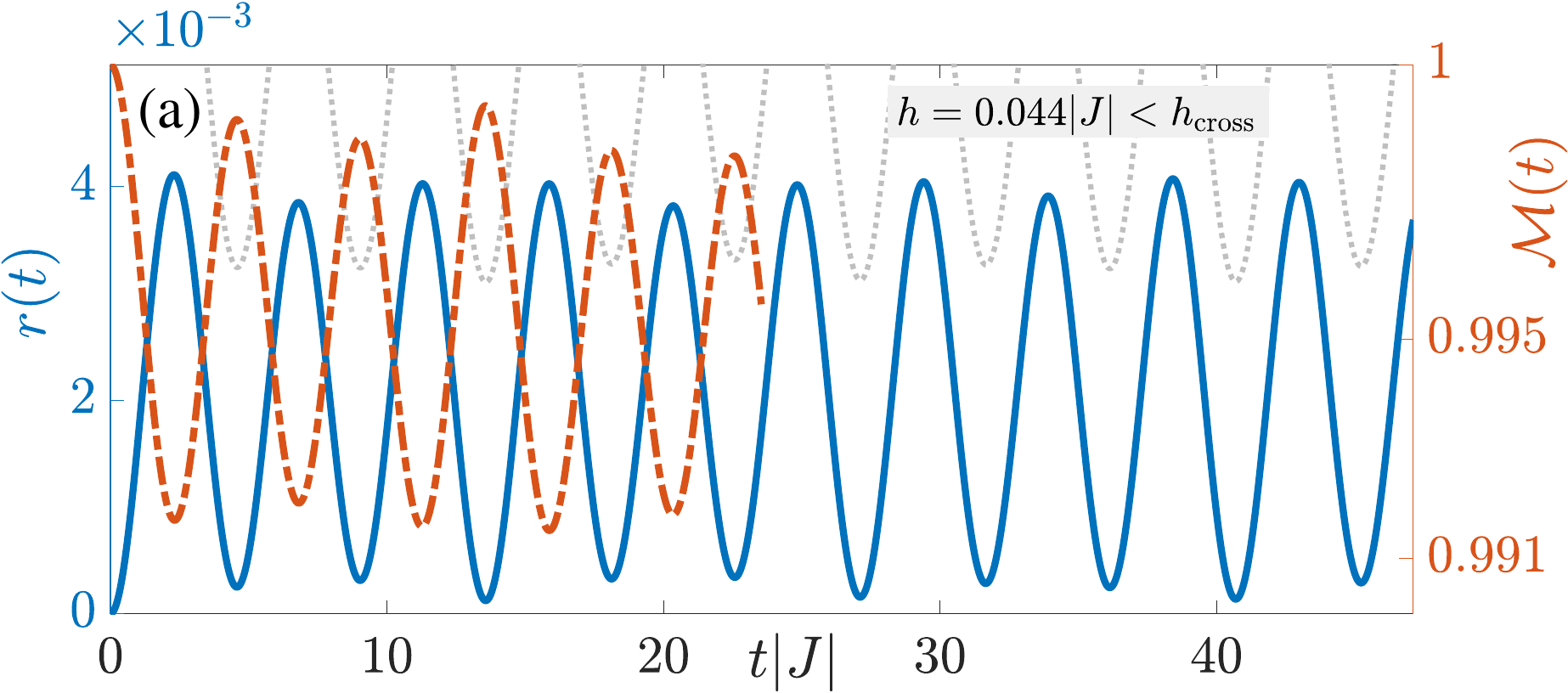}\\
	\includegraphics[width=.48\textwidth]{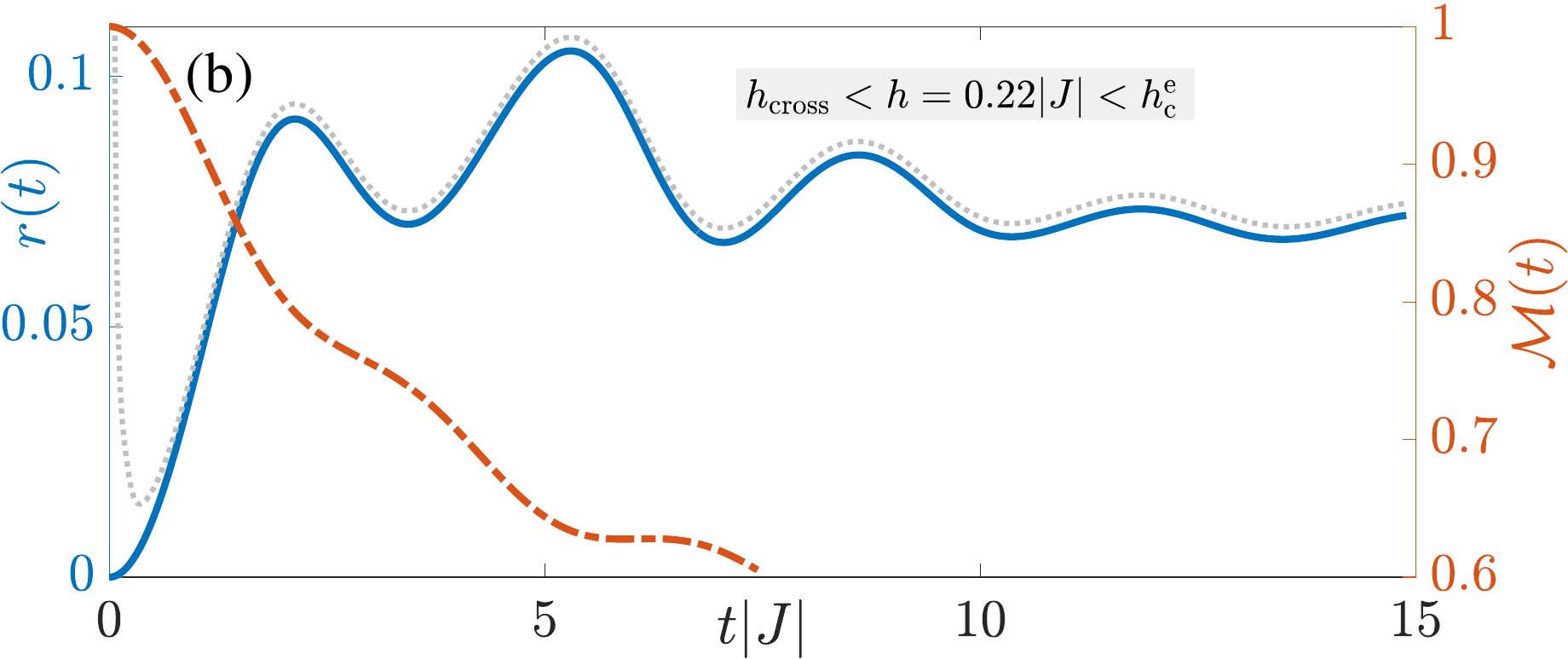}\\
	\includegraphics[width=.48\textwidth]{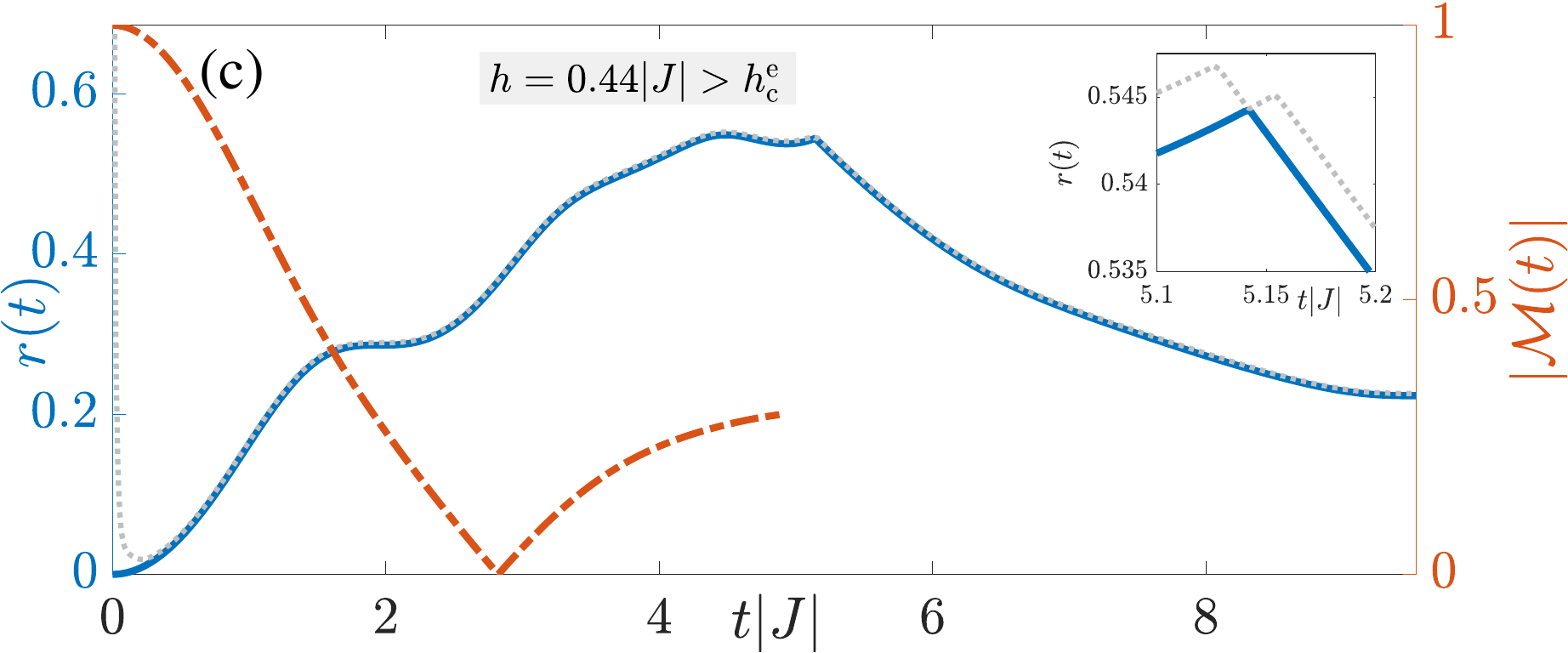}
	\caption{(Color online). Quench dynamics of the Loschmidt-echo return rate (solid blue line), first rate-function branch (dotted gray line; see Appendix~\ref{sec:branch}), and local order parameter (dot-dashed orange line) for the transverse-field Ising chain with antiferromagnetic long-range interactions $\propto1/r^\alpha$ with $\alpha=1.6$ after starting in a N\'eel; see Eq.~\eqref{eq:H} for $J<0$. (a) Unlike the case of ferromagnetic interactions\cite{Halimeh2017} ($J>0$), here anomalous cusps are absent for all evolution times accessible in our iMPS calculations when $h<h_\mathrm{cross}\approx0.051|J|$. Nevertheless, like in the FM case, the order parameter shows persistent oscillations with little decay\cite{Halimeh2017b,Liu2019} at the same frequency as the return rate. (b) Similarly to the ferromagnetic case,\cite{Halimeh2018a} however, cusps are absent when $h_\mathrm{cross}<h<h_\mathrm{c}^\mathrm{d}\approx0.284|J|$, and (c) dynamical criticality appears only in the form of regular cusps when $h>h_\mathrm{c}^\mathrm{d}$, with these cusps coinciding with a zero crossing of the order parameter. Note that within the precision of our iMPS simulations, we have found in the case of antiferromagnetic interactions that the dynamical and equilibrium critical points seem to coincide ($h_\mathrm{c}^\mathrm{d}\approx h_\mathrm{c}^\mathrm{e}$) regardless of $\alpha$, which is different from the case of ferromagnetic interactions\cite{Halimeh2017} where $h_\mathrm{c}^\mathrm{d}\leq h_\mathrm{c}^\mathrm{e}$ with the two critical points coinciding only in the nearest-neighbor limit $\alpha\to\infty$.}
	\label{fig:Fig2}
\end{figure}

Mean-field analysis shows that long-range interactions in the AF-TFIC are relevant for $\alpha\leq2.25$.\cite{Koffel2012} As such, we show in Fig.~\ref{fig:Fig2} the dynamics for $\alpha=1.6$. We consider three different ranges in the values of the final transverse-field strength $h$. The first is $h\in\big(0,h_\mathrm{cross}\big)$, where in the FM phase the return rate can give rise to \textit{anomalous} cusps without any zero crossings in the order parameter.\cite{Halimeh2017,Halimeh2018a} In Fig.~\ref{fig:Fig2}(a), we show the dynamics of the return rate and order parameter at $h=0.044|J|<h_\mathrm{cross}(\alpha=1.6)\approx0.051|J|$. The return rate is smooth without any cusps for all evolution times we are able to access in iMPS even though $h<h_\mathrm{cross}$. This can be explained by noting that in the case of AF interactions, $h_\mathrm{cross}$ is considerably smaller than its counterpart in the FM case for the same value of $\alpha$. This means that anomalous cusps could potentially appear at extremely long evolution times in the AF case, making them inaccessible in iMPS and also modern ion-trap setups. Indeed, similar values of $h$ in the FM case have yielded no anomalous cusps within accessible evolution times, even when larger values $h<h_\text{cross}$ have due to faster dynamics.\cite{Halimeh2017} A spectral analysis of the return rate and the first rate-function branch\cite{Zauner2017} (see Appendix~\ref{sec:branch}) above it in Fig.~\ref{fig:Fig2}(a) reveals that they oscillate at the same frequency within the precision of our numerics, and therefore, we cannot ascertain whether they will intersect at later times to yield nonanalyticities in the return rate. Instead, we see in Fig.~\ref{fig:Fig2}(a)---as we have also checked and found for various values of  $h\in\big(0,h_\mathrm{cross}\big)$---that no anomalous cusps arise even after many cycles of the return rate. It is interesting to note, however, how the dynamics of the order parameter is nevertheless quite constrained, showing little decay over a significant temporal interval, while displaying persistent oscillations at roughly the same frequency as the return rate. This is qualitatively similar to the case of the FM-TFIC for small quenches within the ordered phase.\cite{Halimeh2017b} This is expected from the point of view of perturbation theory: The only eigenstates of the final Hamiltonian that the initial state has significant overlap with, are those with very few single spin flips. At low energies the spectrum of these excitations is roughly linear, fixing the oscillation period, while the minimal magnetization is limited by the number of spin flips in the state with the highest overlap.

The dynamical critical point $h_\mathrm{c}^\mathrm{d}(h_\mathrm{i},\alpha)$ is the final quench value of the transverse-field strength above which the order parameter crosses zero during the time evolution.\cite{Halimeh2018a} In models that host a finite-temperature phase transition, this corresponds to $h_\mathrm{c}^\mathrm{d}$ separating a long-time ordered steady state ($h<h_\text{c}^\text{d}$) from a paramagnetic one ($h>h_\text{c}^\text{d}$). In various models, this has been shown to coincide with a transition in the return rate from exhibiting anomalous cusps or no cusps at all for  $h<h_\mathrm{c}^\mathrm{d}$ to regular cusps for $h>h_\mathrm{c}^\mathrm{d}$.\cite{Heyl2014,Halimeh2018a,Homrighausen2017,Lang2017,Lang2018} This picture persists in the AF-TFIC as well, where in Fig.~\ref{fig:Fig2}(a,b) we see no cusps when $h<h_\mathrm{c}^\mathrm{d}$ and the order parameter does not cross zero within the evolution times accessible in iMPS. However, unlike in Fig.~\ref{fig:Fig2}(a), the dynamics in Fig.~\ref{fig:Fig2}(b) is not constrained. The return rate seems close to equilibration already during the accessible evolution times in iMPS, while the order parameter shows a relatively fast decay. 

The picture qualitatively changes again in Fig.~\ref{fig:Fig2}(c) where $h>h_\mathrm{c}^\mathrm{d}$, and a regular cusp appears during the evolution times obtained in iMPS. As expected,\cite{Heyl2014} this regular cusp connects to a zero crossing of the order parameter. Within the accuracy of our iMPS calculations, we find that for a given value of $\alpha$, the dynamical and equilibrium critical points seem to be roughly the same, $h_\mathrm{c}^\mathrm{d}(h_\mathrm{i},\alpha)\approx h_\mathrm{c}^\mathrm{e}(\alpha)$, with the dynamical critical point showing little dependence on $h_\mathrm{i}$. This is quite different from the case of FM-TFIC, where $h_\mathrm{c}^\mathrm{d}(h_\mathrm{i},\alpha)\leq h_\mathrm{c}^\mathrm{e}(\alpha)$ for $h_\mathrm{i}<h_\mathrm{c}^\mathrm{e}$, with their separation $\lvert h_\mathrm{c}^\mathrm{d}(h_\mathrm{i},\alpha)-h_\mathrm{c}^\mathrm{e}(\alpha)\rvert$ growing larger the smaller $\alpha$ is.\cite{Halimeh2017} In both the FM-TFIC and AF-TFIC, $h_\mathrm{c}^\mathrm{d}(h_\mathrm{i},\alpha)= h_\mathrm{c}^\mathrm{e}(\alpha)$ for  $h_\mathrm{i}>h_\mathrm{c}^\mathrm{e}$, although in this work we do not show results for this case since both models yield the same qualitative picture already discussed for the FM-TFIC in Ref.~\onlinecite{Halimeh2017}.

\begin{figure}[htp]
	\centering
	\includegraphics[width=.48\textwidth]{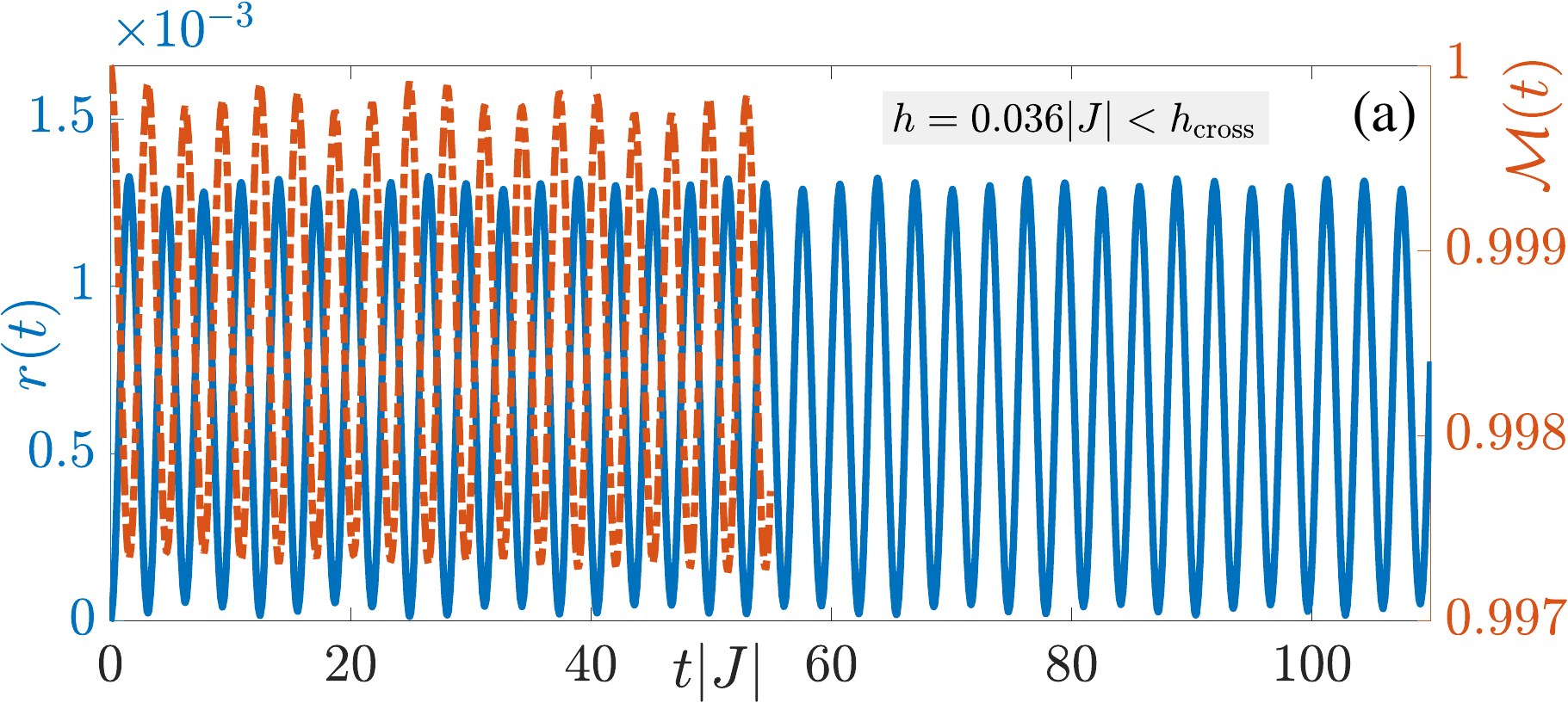}\\
	\includegraphics[width=.48\textwidth]{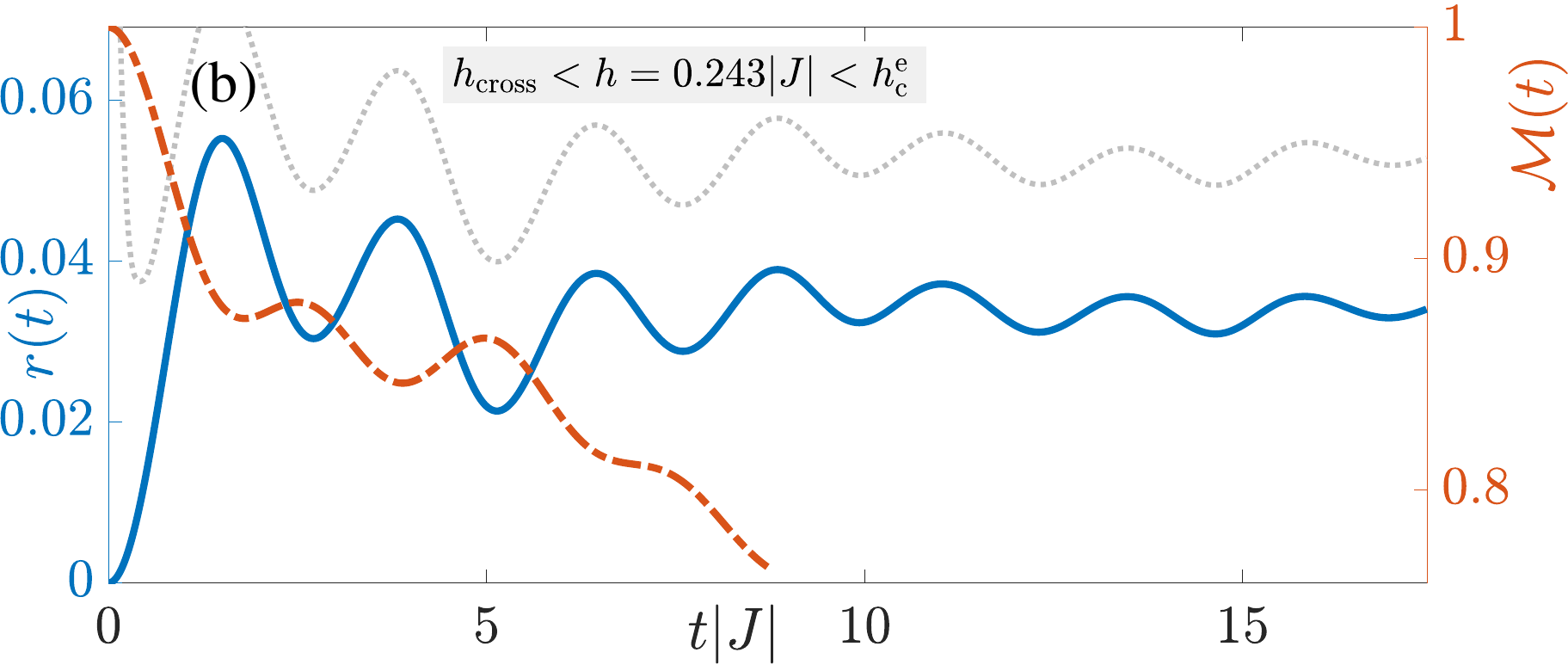}\\
	\includegraphics[width=.48\textwidth]{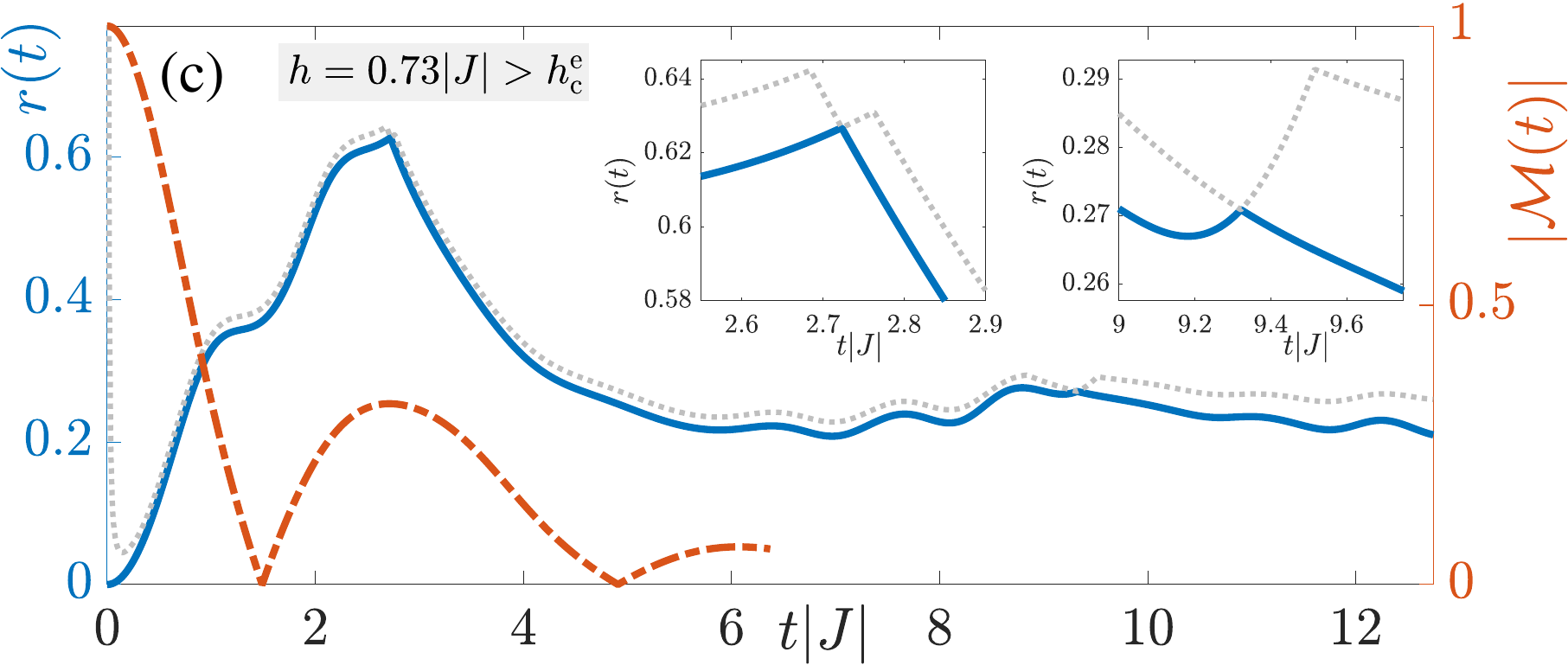}
	\caption{(Color online). Same as Fig.~\ref{fig:Fig2} but for $\alpha=2$. The conclusions are identical with no cusps for $h<h_\mathrm{c}^\mathrm{e}$, and where the return rate exhibits regular cusps corresponding to zero crossings of the order parameter when $h>h_\mathrm{c}^\mathrm{e}$.}
	\label{fig:Fig3}
\end{figure}

For completeness, we repeat these results at $\alpha=2$ in Fig.~\ref{fig:Fig3}, and we arrive at the same conclusions as those from Fig.~\ref{fig:Fig2} for $\alpha=1.6$. The larger value of $\alpha=2$ allows us to push further in evolution times (see Appendix~\ref{sec:LR}), but nevertheless no cusps can be seen for $h<h_\mathrm{c}^\mathrm{e}$ as shown in Fig.~\ref{fig:Fig3}(a,b), and the order parameter is always positive for the evolution times we can access in iMPS. Again worth noting is the slow decay in the order parameter for the small quench $h<h_\text{cross}$ of Fig.~\ref{fig:Fig3}(a), which exhibits persistent oscillations at the same frequency of the return rate. Once again, we have performed a spectral analysis of the return rate and first rate-function branch above it for this case, and within our numerical precision, it seems they oscillate at the same frequency within the accessible evolution times. However, we cannot rule out that at later times they may incur a phase between them that will lead to anomalous cusps at very long times.  Once again, when $h\in(h_\text{cross},h_\mathrm{c}^\mathrm{e})$, we find quick equilibration of the return rate and a relatively fast decay of the order parameter, indicating the absence of constrained dynamics; cf.~Fig.~\ref{fig:Fig3}(b). On the other hand, regular cusps appear when $h>h_\mathrm{c}^\mathrm{e}$, as shown in Fig.~\ref{fig:Fig3}(c). Also in this case the regular cusps correspond to zero crossings in the order parameter.

As a summary of the results presented in this section, the dynamics of the AF-TFIC is fundamentally different from that of its FM counterpart in that small quenches in the ordered phase to $h<h_\text{cross}$ do not yield anomalous cusps during attainable evolution times in iMPS, and therefore, also times currently accessible in modern trapped-ion experiments. This can be understood by noting the much weaker domain-wall binding exhibited in this model compared to the FM-TFIC at the same value of $\alpha$. Furthermore, whereas both dipole and monopole excitations energetically favor smaller domains in the FM-TFIC and can shrink at first-order processes in the transverse-field strength, in the AF-TFIC monopole excitations with larger domains are energetically favored. As such, in the single-domain picture of the AF-TFIC, dipole excitations require at least second-order processes in the transverse-field strength to shrink, and already $h_\text{cross}$ is very small. This weak domain-wall binding behavior seems to at least significantly delay the appearance of anomalous cusps to times currently inaccessible to both numerics and trapped-ion experiments.

\subsection{Integrable case of $\alpha=0$}
For $\alpha>0$, our initial state is a N\'eel state, i.e., a spin configuration with alternating polarization. In a zero-dimensional system with infinite-range interactions, such a state is not unique. A system of $N$ spin-$1/2$ degrees of freedom has $2^N$ eigenstates, which in the case of a system with equal all-to-all interactions separate into only $\mathcal{O}(N^2)$ distinct eigenvalues. This implies that there are large equivalency classes of states that behave indistinguishably during and after a quench. Consequently, it is sufficient to initialize the system in a state that is equivalent to the N\'eel state, which can be expressed in terms of total spin length $S$ and magnetization $m$ along the $z$-axis.
We detail the derivation in Appendix~\ref{sec:Neel}, which gives the normalized result
\begin{align}\label{Neel}
\ket{\psi_0}=\sum_S\sqrt{\frac{D(S)}{\sum_{S'}D(S')}}\ket{S,m=1/2},
\end{align}
where the degeneracy of the subspace with spin length $S$ is
\begin{align}\label{eq:degen}
D_N(S)=\frac{2S+1}{N+1}{N+1 \choose \frac{N}{2}-S}.
\end{align}
Within each $\ket{S,m}$ subspace, there is only trivial dynamics, which thus does not need to be resolved.

Following the initialization in a N\'eel-equivalent state, the system is evolved in time with the Lipkin-Meshkov-Glick (LMG) Hamiltonian.\cite{Lipkin1965} The latter is obtained, up to a trivial overall factor, in the limit of infinite-range interactions ($\alpha=0$) of Eq.~\eqref{eq:H} and expressed in terms of collective spin operators $\hat{S}^a=\frac{1}{2}\sum_{i=1}^N\sigma_i^a$ as
\begin{align}\label{eq:H_LMG}
H=-\frac{J}{2N}\hat{S}^z\hat{S}^z-\Gamma \hat{S}^x,
\end{align}
with $\Gamma=h/4$, in keeping with the LMG notation.\cite{Lipkin1965}

In the case of FM interactions ($J>0$), the LMG model has a second-order phase transition at the quantum critical point $\Gamma_\text{c}^\text{e}=J/2$,\cite{Botet1982,Botet1983} with a corresponding dynamical critical point $\Gamma_\text{c}^\text{d}=(\Gamma_\text{i}+\Gamma_\text{c}^\text{e})/2$, for quenches starting at $\Gamma_\text{i}\leq\Gamma_\text{c}^\text{e}$, and $\Gamma_\text{c}^\text{d}=\Gamma_\text{c}^\text{e}$, for quenches starting at $\Gamma_\text{i}\geq\Gamma_\text{c}^\text{e}$.\cite{Homrighausen2017,Lang2017} In the case of AF interactions ($J<0$), the LMG model exhibits a first-order phase transition at the quantum critical point $\Gamma_\text{c}^\text{e}=0$.\cite{Vidal2004,Chen2009}

We now focus on the case of AF interactions and, without loss of generality, we set in the following $J=-1$. The conservation of total spin length $S$ for $\alpha=0$ has a major impact on the return rate. In case of a thermal initial state, it has been argued that the (interferometric) return rate employed here is unsuitable for the description of dynamical phase transitions.\cite{Lang2017,Sedlmayr2018} Instead, the Loschmidt amplitude should be replaced by the fidelity of the initial and final density matrix, or similarly by summing all Loschmidt amplitudes within the subspaces of fixed spin.\cite{Lang2017} However, for the present case of a pure initial state, the situation is quite different. The fidelity return rate coincides with the definition~\eqref{eq:ReturnRate}, but the geometric connection with the quantum return rate is lost as the corresponding wave packet on the Bloch sphere is extremely broad, thereby breaking the semiclassical picture that connected the return rates in the first place.\cite{Lang2018} It is, however, possible to recover a geometric interpretation as follows. Consider the N\'eel-equivalent state with all up-spins sorted to the left,
\begin{align}\label{eq:Neel}
\ket{\psi_0}=\ket{\uparrow \dots \uparrow \downarrow \dots \downarrow}=\ket{\frac{N}{4},\frac{N}{4}}\ket{\frac{N}{4},-\frac{N}{4}}\,.
\end{align}
The collective spin operators can be split into operators acting exclusively on either the left or the right part of the system ($S^\alpha=S^\alpha_l+S^\alpha_r$) such that the Hamiltonian \eqref{eq:H_LMG} describes two coupled LMG models, each with half the original spin length
\begin{align}\label{eq:Hsplit}
H=\sum_{s=l,r}\left(\frac{1}{2N}\hat{S}^z_s\hat{S}^z_s-\Gamma \hat{S}^x_s\right)+\frac{1}{N}\hat{S}^z_l\hat{S}^z_r\,.
\end{align}
Parametrizing the classical spin vector $\mathbf{S}=S\left(\sin{\theta}\cos{\phi},\sin{\theta}\sin{\phi},\cos{\theta}\right)^\intercal$, we obtain the classical equations of motion in the form
\begin{subequations}
\begin{align}
\frac{d \theta_s}{d t}&=-\Gamma \sin{\phi_s}\\
\frac{d \phi_s}{d t}&=-\Gamma\cos{\phi_s}\cot{\theta_s}+\frac{1}{4}\sum_{\eta=l,r}\cos{\theta_\eta}\,,
\end{align}
\end{subequations}
with $s\in\{l,r\}$, from which we can immediately read off the modulus of the staggered magnetization in the thermodynamic limit as
\begin{align}\label{eq:M}
\mathcal{M}(t)=\lvert\cos(\Gamma t)\rvert\,.
\end{align}
The advantage of the representation via two coupled systems lies in the well-localized spin WKB wave function \cite{VanHemmen1986,VanHemmen2003,Lang2018} of the initial state~\eqref{eq:Neel}, for which furthermore all return rates coincide. It is now straightforward to apply the geometric interpretation\cite{Lang2018} to obtain
\begin{align}\label{eq:r_WKB}
r(t)\approx \min_{\theta,\phi}\frac{1}{2}\sum_{s=l,r}\frac{1}{4}\left[\sin^2{\vartheta_s(\theta_l,\theta_r,\phi_l,\phi_r|t)}+\sin^2{\theta_s}\right]\,,
\end{align}
where $\vartheta_s(\theta_l,\theta_r,\phi_l,\phi_r|t)$ is the time-evolved polar angle with initial coordinates $\theta_l,\theta_r,\phi_l,\phi_r$.

\begin{figure}[htp]
	\centering
	\includegraphics[width=.48\textwidth]{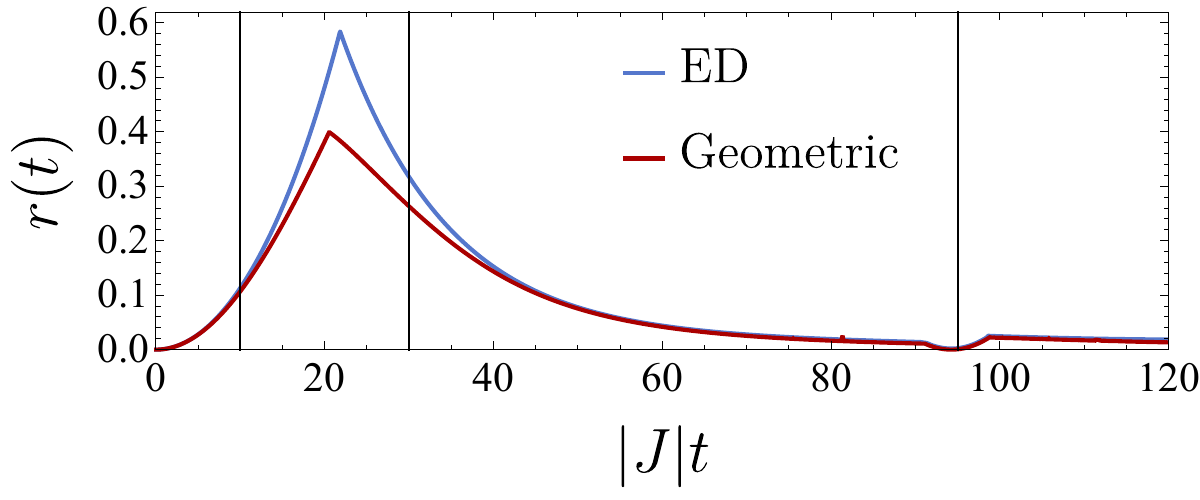}
	\caption{(Color online). Return rate in the LMG model given in Eq.~\eqref{eq:H_LMG} after starting in the initial state~\eqref{eq:Neel} and quenching to $\Gamma=\lvert J\rvert/15$. ED result (blue) is obtained for $N=1201$ spins. Geometric result is shown in red. Regardless of the value of the latter, the return rate exhibits regular cusps. The vertical lines indicate the snapshots in evolution time $t\lvert J\rvert=10,\,30,\,95$ at which the corresponding Bloch sphere representation is shown in Fig.~\ref{fig:Fig6} from left to right, respectively.}
	\label{fig:Fig5}
\end{figure}

\begin{figure*}[htp]
	\centering
	\includegraphics[width=.32\textwidth]{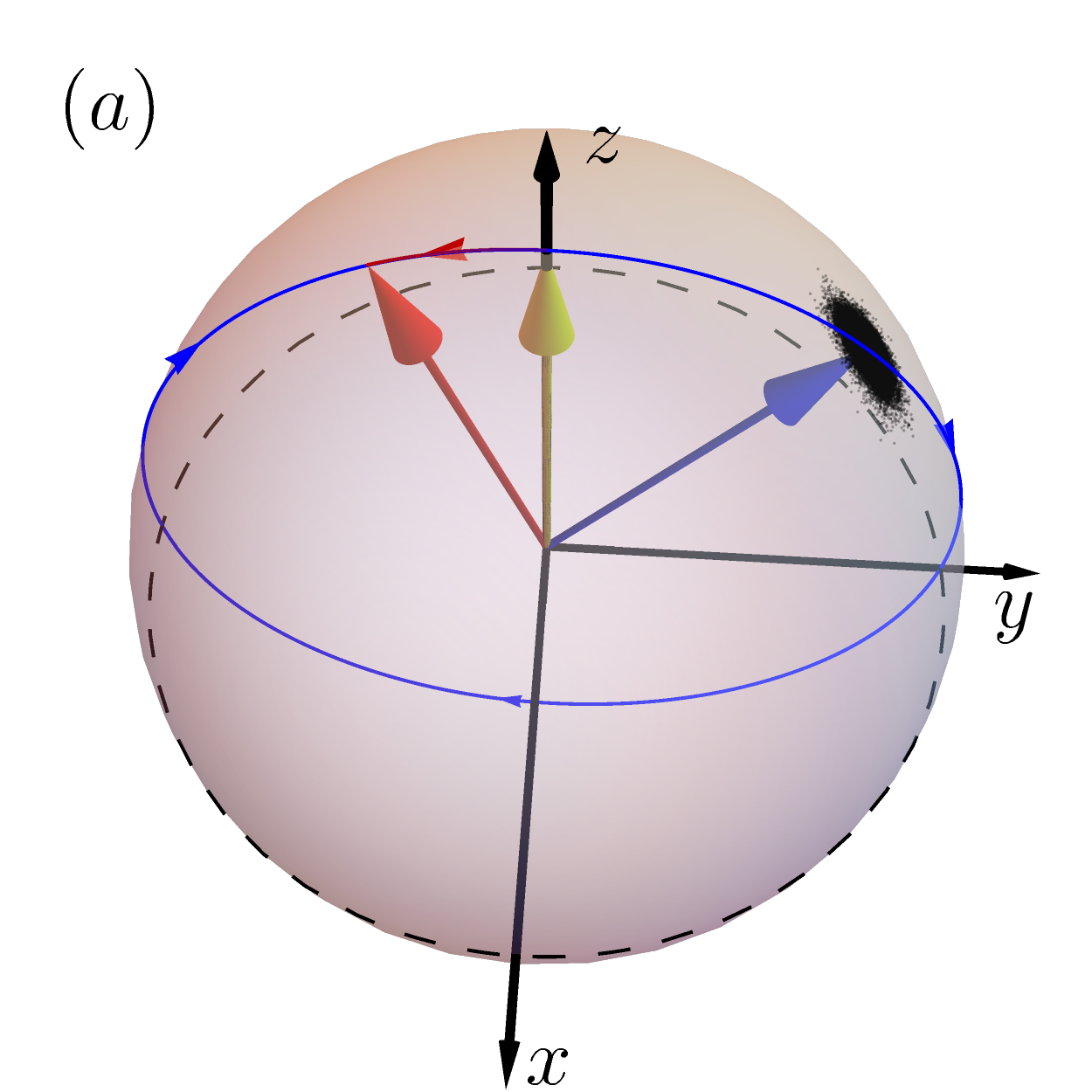}\includegraphics[width=.32\textwidth]{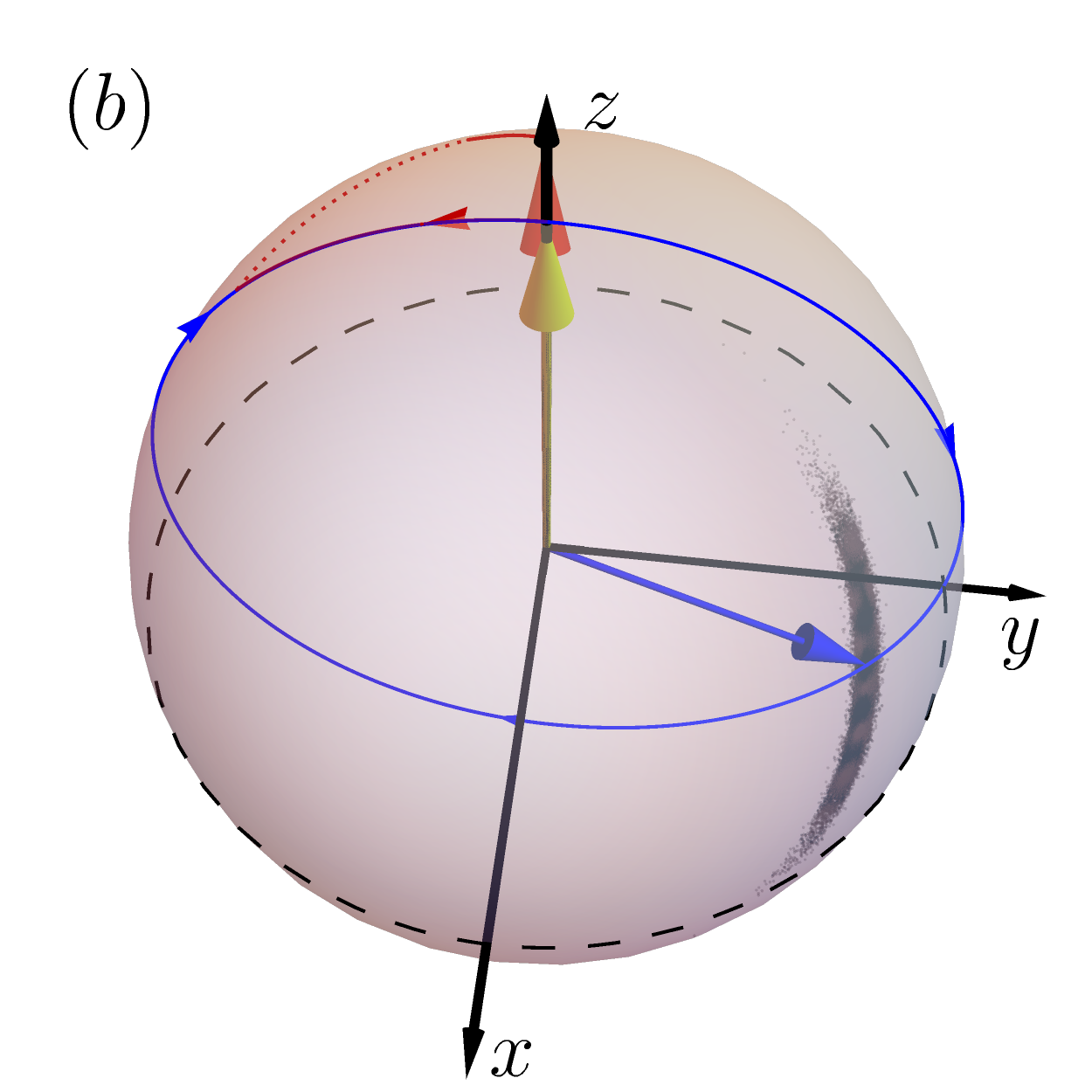}\includegraphics[width=.32\textwidth]{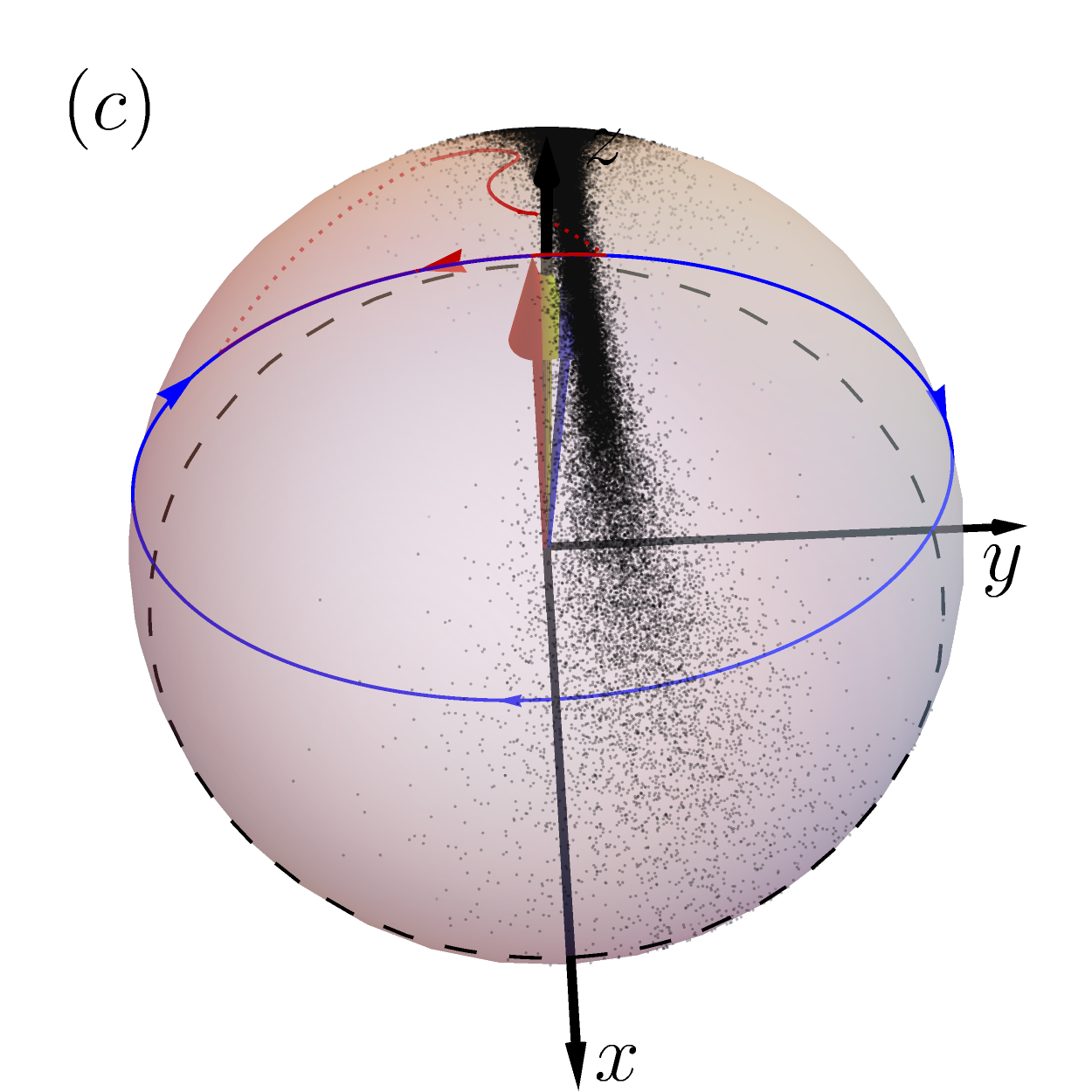}\\
	\includegraphics[width=.32\textwidth]{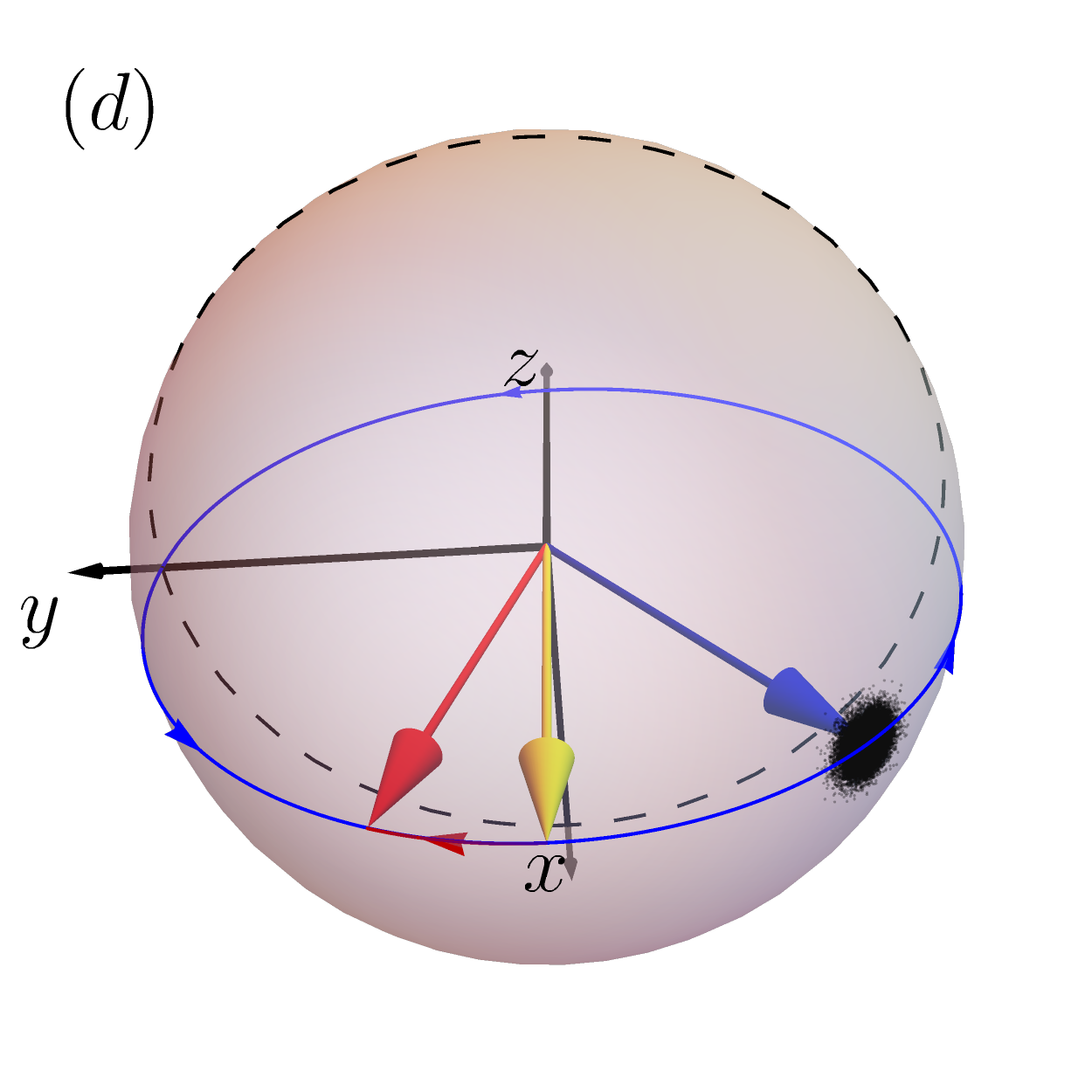}\includegraphics[width=.32\textwidth]{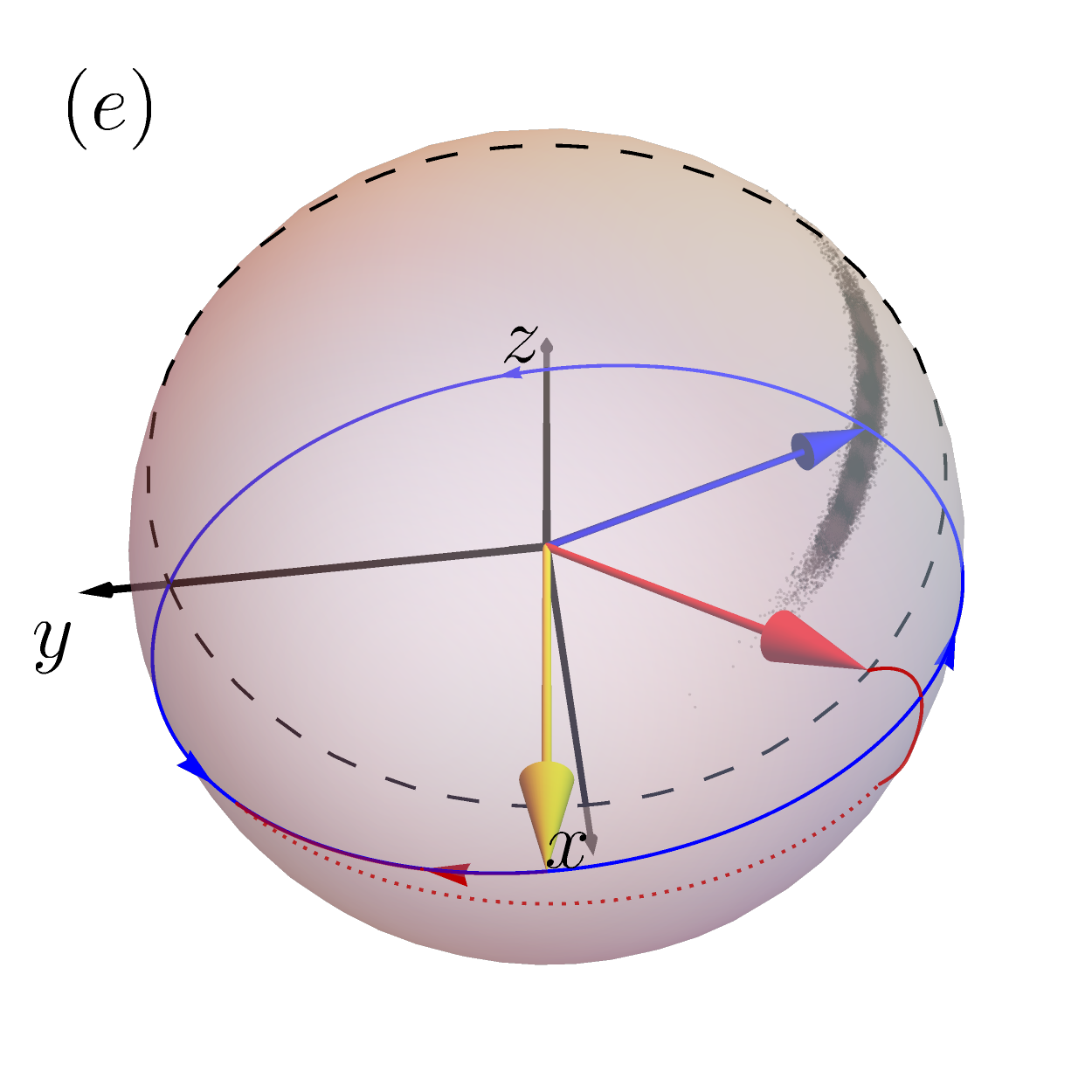}\includegraphics[width=.32\textwidth]{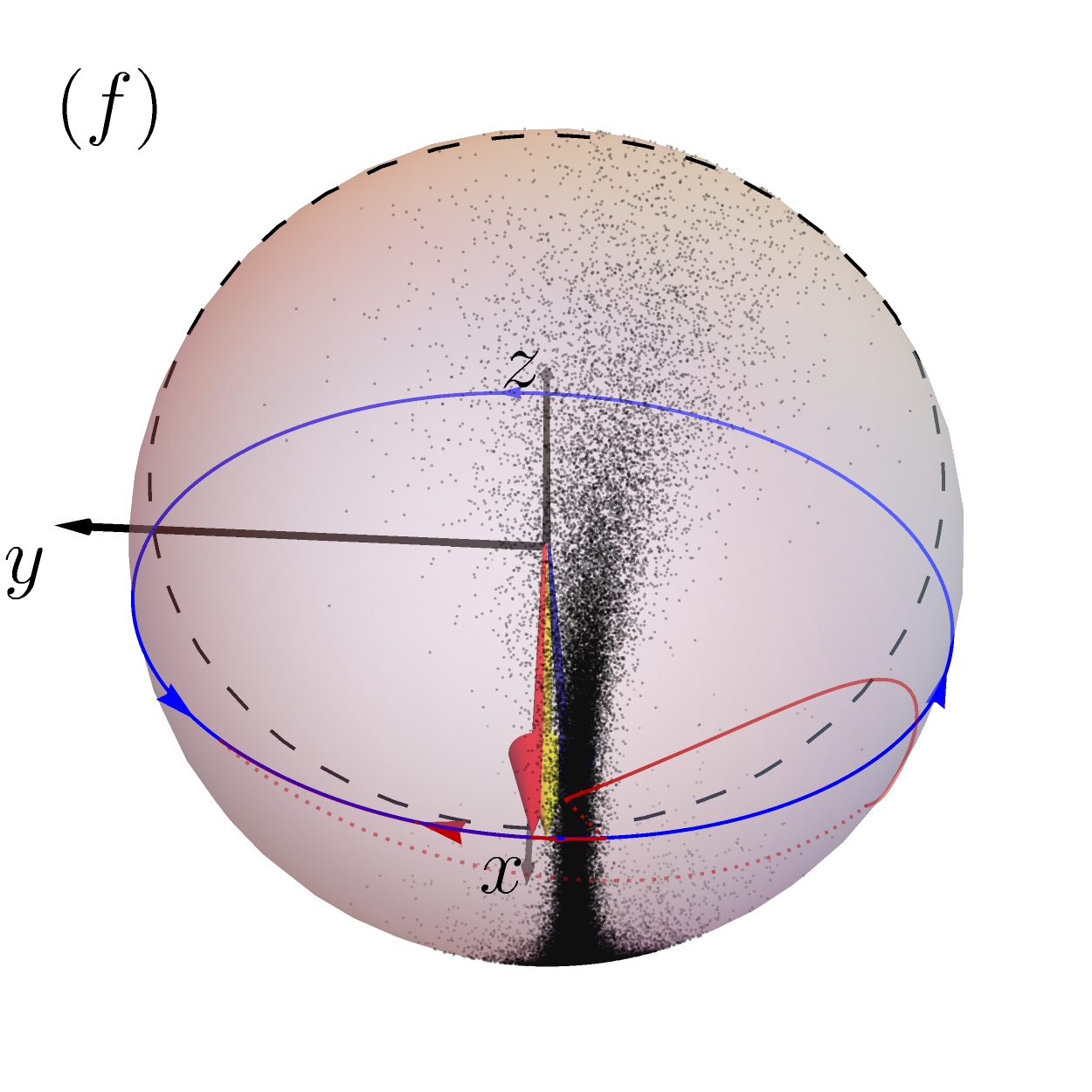}
	\caption{(Color online). Bloch sphere representation of the spin WKB wave function (shown as a flock of initialization points) together with the trajectories of the magnetization (blue) and the Loschmidt vector (red). Dashed black circles indicate the equator. Depicted is a quench from the initial state~\eqref{eq:Neel} to $\Gamma=\lvert J\rvert/15$ at times (a,d) $t=10/\lvert J\rvert$, (b,e) $t=30/|J|$, and (c,f) $t=95/|J|$ for the (a,b,c) left and (d,e,f) right subsystems of the WKB wave function. At each time the current magnetization and Loschmidt vector are highlighted as blue and red vectors respectively. The top row shows the part of the system initially aligned parallel to the $z$-axis, while the bottom row displays the antiparallel half with the initial orientation of both marked in yellow. The wave function spreads quickly as fluctuations are enhanced compared to its ferromagnetic counterpart.\cite{Lang2018} Interestingly, after an odd number of cusps the Loschmidt echo is dominated by states with finite magnetization, despite starting in an unpolarized initial state.}
	\label{fig:Fig6}
\end{figure*}

For the LMG model, every quench evolving the N\'eel state with $\Gamma\neq 0$ results in a regular return rate with two cusps in every period, as shown in Fig.~\ref{fig:Fig5}. This behavior is well captured by the spin WKB Ansatz, which shows remarkable agreement with the corresponding ED result in the estimated critical times. Note, however that the height of the first peak is significantly underestimated since in the steep tails of the wave function gradient corrections become increasingly important. We also show in Fig.~\ref{fig:Fig6} the WKB wave function on the Bloch sphere of the left subsystem (top) and its right counterpart (bottom), corresponding to evolution times marked by vertical lines in Fig.~\ref{fig:Fig5}. It is interesting to note that during the early evolution shown in Fig.~\ref{fig:Fig6}(a,d) the Loschmidt vector, i.e., the time-evolved saddle-point coordinates of Eq.~\eqref{eq:r_WKB}, evolves in the opposite direction of the magnetization vector, exemplifying the enhanced spreading of the AF WKB wave function. At the first cusp, the Loschmidt vector spontaneously breaks the mirror symmetry $x\to-x$; see Fig.~\ref{fig:Fig6}(b,e). As depicted in Fig.~\ref{fig:Fig6}(c,f), only shortly before the magnetization returns to its initial value does the symmetric saddle point take over again, resulting in two cusps per period as also observed in the ED result of Fig.~\ref{fig:Fig5}.

Phenomenologically, it is not surprising that the return rate in the case of the LMG model with AF interactions only regular cusps appear after quenching from the N\'eel state, since at $T=0$ the AF LMG has no order at any $\Gamma>\Gamma_\text{c}^\text{e}=0$ . As such, any quenches from a N\'eel state will lead to regular cusps in the return rate, and correspondingly an order parameter oscillating around zero, in accordance with Eq.~\eqref{eq:M}.

\section{Experimental considerations}
\label{sec:experimental}
In this section, we discuss the observability of our results in ion-trap setups. In this platform, spins-$1/2$ are represented by a suitable choice of two internal electronic states, between which interactions are transmitted via a phonon bus using lasers or microwave fields that are tuned close to the vibrational modes of the ion crystal,\cite{Porras2004,Zippilli2014} e.g., using a Moelmer--Soerensen configuration.\cite{Moelmer2000}
	
An experimental sequence to study the cusps in the AF-TFIC may proceed as follows:
	(i) First, the system is initialized in a N\'eel product state, which can be done with near-perfect precision using optical pumping.\cite{Blatt2012,Schindler2013,Monroe2021} 
	(ii) Then, the electromagnetic fields are turned on that generate the effective spin--spin Hamiltonian. 
	(iii) After an adjustable evolution time, the Loschmidt amplitude can be read out by projecting onto the initial state.\cite{Jurcevic2017} Since this is a product state, standard fluorescence imaging of the individual electronic states of the ions is sufficient, which again can be performed with high precision. \cite{Blatt2012,Schindler2013,Monroe2021} 
	Although observing the overlap with a specific state is exponentially inefficient, such measurements have been demonstrated for systems up to $N=16$,\cite{Islam2013} and in special cases even for more than $N=100$ ions.\cite{Gaerttner2017} 
	In the context of DQPTs, the projectors have been measured for up to $N=10$ ions in a ferromagnetic setting, from which cusps at $N\to\infty$ have been reliably extrapolated.\cite{Jurcevic2017} 

The Hamiltonian governing step (ii) is worthy of some further discussion. The natural spin--spin interactions in trapped ions are---due to the collective character of the phonon modes---long ranged, with a decrease over distance that can be adjusted by tuning the frequency of the Moelmer--Soerensen fields relative to the phonon frequencies.\cite{Porras2004} In many experiments, these interactions are well described by a spatial power-law decay,  \cite{Gaerttner2017,Islam2013,Richerme2014,Jurcevic2014} an approximation that works rather well for small systems.\cite{Nevado2016,Trautmann2018} 
Increasing the detuning to the phonon frequencies, larger power-law exponents can be achieved, with a range of $\alpha$ accessible with current laser powers that lies approximately between $0.6$ and $1.5$.\cite{Jurcevic2014} In addition, by tuning close to the lowest mode, which has the largest frequency separation from  the other modes, $\alpha=0$ can be achieved. 
In current experiments, the power-law interactions are induced using the radial phonon modes, which are closer spaced than the axial modes.\cite{Porras2004} In such a configuration, the lowest-frequency mode is the zig-zag mode,\cite{Lechner2016} which has opposite oscillation amplitude at neighboring ions. This pattern is imprinted on the generated effective interactions, which as a consequence are naturally of antiferromagnetic type. 
There have also been various proposals to precisely engineer desired interaction patterns using several laser fields \cite{Korenblit2012,Hauke2015,Davoudi2020} or segmented Paul traps,\cite{Zippilli2014} which could overcome any imprecision in the power-law couplings and extend the range of accessible power-law exponents. 
As a final note, a transverse field can easily be implemented in the Moelmer--Soerensen configuration using an asymmetric detuning to the blue and red phonon sidebands.\cite{Jurcevic2014}

In such experiments, a main potential restriction for the observability of the cusps is the achievable evolution times, which are limited due to dissipative processes such as spontaneous emission or fluctuations of magnetic fields or laser intensities. Nevertheless, current experiments reach evolution times on the order of several $\mathcal{K}_\alpha/J$,\cite{Richerme2014,Jurcevic2014,Neyenhuis2017,Gaerttner2017,Jurcevic2017,Gaerttner2017} which are on the order of the times studied here numerically. Hence, the physics discussed here on the basis of numerical simulation should be observable in state-of-the-art trapped-ion experiments.

\section{Conclusions and outlook}\label{sec:conclusion}
We have investigated dynamical phase transitions in transverse-field Ising chains with antiferromagnetic power-law-decaying interactions through exact diagonalization and infinite matrix product states. Although the behavior is similar to the ferromagnetic case when interactions are short-range, in the long-range case the behavior is qualitatively different. Whereas anomalous cusps are a prominent feature of quenches within the ordered phase in the case of long-range ferromagnetic interactions, they are absent during all accessible evolution times in the case of even the longest-range antiferromagnetic interactions that we can numerically achieve. We attribute this to the significantly weaker overall domain-wall binding in the antiferromagnetic case, and additionally the fact that monopole excitations are repulsive unlike in the ferromagnetic case where both dipoles and monopoles are energetically favored to bind for small quenches. 

This picture persists even in the integrable case of infinite-range interactions. There, domain walls are not defined even in the ordered phase, and local spin flips (equivalent to bound domain walls in $1$D) are the lowest-lying quasiparticles. However, in this case the model hosts a first-order phase transition at zero transverse-field strength in equilibrium at zero temperature. Consequently, quenching from the ground-state manifold will always lead to regular cusps and concomitantly an order parameter oscillating around its long-time steady-state value of zero. 

It is important to note that we cannot rule out anomalous cusps arising at very long times for small quenches within the ordered phase in the regime with domain-wall binding in the antiferromagnetic case. However, it seems certain that these times are inaccessible in our numerics and, therefore, are likely too long to be reached in modern quantum synthetic matter setups.

Our work further highlights the crucial role constrained dynamics plays in dynamical phase transitions, and how long-range interactions are neither a sufficient nor necessary condition for anomalous dynamical criticality. Indeed, we find no anomalous cusps for all accessible evolution times in the antiferromagnetic case even when long-range interactions are relevant (exponent $\alpha\leq2.25$). However, it is known that quantum Ising chains with ferromagnetic exponentially decaying interactions, which are in the short-range universality class in equilibrium, will still give rise to anomalous cusps because one can achieve a profile there where domain-wall binding is prominent.\cite{Halimeh2018a} Our findings therefore provide further insight into the nature of anomalous cusps and under what conditions they are prominent in the return rate.

The results presented in this paper can be observed in modern ion-trap setups where long-range antiferromagnetic interactions are naturally induced. This is due to the fact that the lowest-lying and best-separated mode of radial phonons is the zig-zag one.\cite{Lechner2016} This should make the detection of dynamical phase transitions in long-range antiferromagnetic quantum Ising chains a natural target following their ferromagnetic counterparts, which have already been investigated.\cite{Jurcevic2017}

\begin{acknowledgments}
We acknowledge  support by Provincia Autonoma di Trento, the ERC Starting Grant StrEnQTh (project ID 804305), Q@TN  —  Quantum  Science  and  Technology  in  Trento — and the Collaborative Research Centre ISOQUANT (project ID 273811115), Research Foundation Flanders (G0E1520N, G0E1820N), and ERC grants QUTE (647905) and ERQUAF (715861).
\end{acknowledgments}
	
\appendix
\section{Two-kink model}\label{sec:kink}
The single-domain wall states in the effective Hilbert space of the two-kink model can be denoted $\ket{j,M}$, where $j$ indicates the spatial position of the first site of the domain of length $M$. As such, the state $\ket{j,M}$ has its spins on sites $j$ up to $j+M-1$ all flipped with respect to a ground state of Eq.~\eqref{eq:H} at zero transverse-field strength. This ground state is doubly degenerate for $\alpha>0$, and is thus one of two possible N\'eel states in the case of AF interactions, or a fully $z$-up or $z$-down polarized state when the interactions are FM; see Fig.~\ref{fig:statics}(a). It is also evident that at zero strength of the transverse field, the states $\ket{j,M}$ are eigenstates of Eq.~\eqref{eq:H}. The two-kink Hamiltonian can thus be written as
\begin{align}\nonumber
	H_\mathrm{kink}=&\sum_j\sum_M \Big\{V_{\alpha,\chi}(M)\ket{j,M}\bra{j,M}\\\nonumber
	&-h\big[\ket{j,M}\big(\bra{j,M-1}+\bra{j+1,M-1}\\
	&+\bra{j,M+1}+\bra{j-1,M+1}\big)+\text{H.c.}\big]\Big\}.
\end{align}
The terms $\propto h$ correspond to the growth or shrinking of the domain via a magnetic-field induced spin-flip. 

\section{Details regarding iMPS calculations}\label{sec:NumSpec}
In this Appendix, we explain how power-law interactions are realized in our iMPS implementation, we discuss the backward time-evolution trick employed to double the maximal time evolution reached in the return rate, and we define rate-function branches from the eigenvalues of the MPS transfer matrix.

\subsection{Implementation of long-range interactions}\label{sec:LR}
In a strict sense, power-law interactions are not possible to implement in iMPS, because the latter is based on matrix product operator (MPO) formulations of exponentials in Hamiltonian parts composed of commuting terms. Naturally, this is ideal for systems where an MPO description is possible such as when interactions are exponentially decaying. However, up to good accuracy, a power-law decay can be approximated by a sum of exponentials, with the number of the latter increasing the longer-range the power-law interactions are. Consequently, a sum of the MPO representations of these exponentials will accurately model the corresponding power-law interaction profile.\cite{Crosswhite2008} The approximation of the power-law profile with exponentials is given by
\begin{align}
	\sum_{j<l}\frac{1}{\lvert j-l\lvert^\alpha}\sigma_j^z\sigma_l^z\approx\sum_{j<l}\sum_{n=1}^Kc_nu_n^{\lvert j-l\lvert-1},
\end{align}
where $K$ is the number of exponentials used. The coefficients $c_n$ and $u_n$ are real numbers with $0\leq u_n<1$, and they are computed using a nonlinear least-squares fit. Our fidelity threshold for this fit is an error $\varepsilon\approx\mathcal{O}(10^{-10}-10^{-8})$, where $K$ is chosen over a distance $d$ such that $\sum_{n=1}^Kc_nu_n^{d-1}<\varepsilon$. Consequently, $K$ increases the smaller $\alpha$ is, thereby limiting how long-range the interactions we implement can be. For more details on the implementation of long-range interactions in iMPS, we refer the reader to Ref.~\onlinecite{Crosswhite2008}.

\subsection{Backward time-evolution trick}\label{sec:backward}
This trick entails noticing that the Loschmidt amplitude can be rewritten as $G(t)=\bra{\psi_0}e^{-iHt}\ket{\psi_0}=\bra{\psi_0}e^{-iHt/2}e^{-iHt/2}\ket{\psi_0}$, which means that to reach an evolution time $t$ in the Loschmidt return rate, one need only time-evolve $\ket{\psi_0}$ up to $t/2$ to obtain $\ket{\psi(t/2)}=\exp(-iHt/2)\ket{\psi_0}$, which is then complex conjugated to obtain $\ket{\psi(-t/2)}=\overline{\ket{\psi(t/2)}}=\exp(iHt/2)\ket{\psi_0}$. In iMPS, we therefore evolve $\ket{\psi_0}$ forward in time to a maximal evolution time $t_\text{max}$ (within convergence at a certain time-step and bond dimension), allowing results for $r(t)$ up to $2t_\text{max}$. Since this trick cannot be applied to the local order parameter $\mathcal{M}(t)$, results for the latter are only available up to $t_\text{max}$.\cite{Zauner2017}

\subsection{MPS transfer matrix and rate-function branches}\label{sec:branch}
According to the definition of the return rate given in Eq.~\eqref{eq:ReturnRate}, the latter is the negative of the natural logarithm of the dominant eigenvalue of the mixed MPS transfer matrix $\mathcal{T}(t)$ between MPS tensors at evolution times $0$ and $t$.\cite{Zauner2017} The spectral radius of $\mathcal{T}(t)$ is bounded above by unity. Let us order the eigenvalues $\epsilon_n(t)$ of $\mathcal{T}(t)$ in descending order by magnitude: $\epsilon_1(t)\geq\epsilon_2(t)\geq\ldots\geq\epsilon_\mathcal{D}(t)$. Then the rate-function branches are
\begin{align}
	r_n(t)=-2\ln\lvert\epsilon_n(t)\rvert.
\end{align}
Consequently, the return rate is $r(t)=r_1(t)\leq r_2(t)\leq r_3(t)\leq\ldots\leq r_\mathcal{D}(t)$. At a critical time $t_\mathrm{c}$, when a cusp appears in the return rate, the return rate and the first rate-function branch are equal, $r(t_\text{c})=r_2(t_\text{c})$. This is particularly useful since the nonanalyticities of DPT-II are nothing but level crossings of the eigenspectrum of $\mathcal{T}(t)$. Such level crossings are characteristic of first-order phase transitions. 

It is worth noting that $\mathcal{T}(t)$ is an approximation of the physical quantum transfer matrix arising in a real-time path-integral formulation of the Loschmidt amplitude $G(t)=\bra{\psi_0}e^{-iHt}\ket{\psi_0}$.

\section{N\'eel state with permutation invariance}\label{sec:Neel}
A N\'eel state has minimal magnetization, thus for an odd number $N$, we can choose (without loss of generality) $m=1/2$. The spin length, on the other hand, is a bit more subtle, since the N\'eel state is no eigenstate of $\hat{S}^2=\hat{S}_x^2+\hat{S}_y^2+\hat{S}_z^2$ with $\hat{S}_a=\frac{1}{2}\sum_{i=1}^N\sigma_i^a$.

To explain the general procedure, let us first consider a system of only three spins with the N\'eel state $\ket{\uparrow\downarrow\uparrow}$, which we want to express in terms of the simultaneous eigenstates of $\hat{S}^2$ and $\hat{S}^z$ denoted as
\begin{subequations}\label{Neel3spins}
\begin{align}
\ket{\frac{3}{2},\frac{1}{2}}&=\frac{1}{\sqrt{3}}\big(\ket{\uparrow\uparrow\downarrow}+\ket{\uparrow\downarrow\uparrow}+\ket{\downarrow\uparrow\uparrow}\big)\\
\ket{\frac{1}{2},\frac{1}{2}}_a&=\sqrt{\frac{2}{3}}\Big(\ket{\uparrow\uparrow\downarrow}-\frac{1}{2}\ket{\uparrow\downarrow\uparrow}-\frac{1}{2}\ket{\downarrow\uparrow\uparrow}\Big)\\
\ket{\frac{1}{2},\frac{1}{2}}_b&=\frac{1}{\sqrt{2}}\big(\ket{\uparrow\downarrow\uparrow}-\ket{\downarrow\uparrow\uparrow}\big).
\end{align}
\end{subequations}
This is achieved by
\begin{align}\label{Neel3spinsfull}
\ket{\uparrow\downarrow\uparrow}=\frac{1}{\sqrt{3}}\ket{\frac{3}{2},\frac{1}{2}}-\frac{1}{\sqrt{6}}\ket{\frac{1}{2},\frac{1}{2}}_a+\frac{1}{\sqrt{2}}\ket{\frac{1}{2},\frac{1}{2}}_b,
\end{align}
which involves the states $\ket{\frac{1}{2},\frac{1}{2}}_a$ and $\ket{\frac{1}{2},\frac{1}{2}}_b$, which are indistinguishable with respect to the time-evolution. For the infinite-ranged system these behave identically, such that we can write the equivalency statement
\begin{align}\label{3spins}
\ket{\uparrow\downarrow\uparrow}=\frac{1}{\sqrt{3}}\ket{\frac{3}{2},\frac{1}{2}}+\sqrt{\frac{2}{3}}\ket{\frac{1}{2},\frac{1}{2}}.
\end{align}
This construction can be extended to arbitrary system sizes with $N$ spins. Since the Hamiltonian is integrable there exist $N$ conserved quantities, i.e., the $N-1$ quantities $\vec{S}^2_n\equiv S^{x2}_{n}+S^{y2}_{n}+S^{z2}_{n}$ ($n=2,\dots,N$) and the Hamiltonian itself satisfy $[\vec{S}^2_n, H_{\alpha=0}]=0$, where $S^{\beta}_{n}=\sum_{i=1}^ns^{\beta}_i$ with $\beta = x,y,z $. Therefore, we can label each eigenstate by $\ket{S_1,S_{2},\cdots,S_{n},\cdots,S_N,S^z_N}$, where $S_n\in \{ |S_{n-1}-\frac{1}{2}|,\cdots,S_{n-1}+\frac{1}{2}\}$. For example, the three states in Eqs.~\eqref{Neel3spins}, $\ket{\frac{3}{2},\frac{1}{2}}$, $\ket{\frac{1}{2},\frac{1}{2}}_a$ and $\ket{\frac{1}{2},\frac{1}{2}}_b$, correspond to $\ket{1,\frac{3}{2},\frac{1}{2}}$, $\ket{1,\frac{1}{2},\frac{1}{2}}$ and $\ket{0,\frac{1}{2},\frac{1}{2}}$ respectively. We can represent the LMG eigenstate $\ket{S_{2},\cdots,S_{n},\cdots,S_N,S^z_N}$ in the uncoupled tensor basis $\ket{s^z_1,s^z_{2},\cdots,s^z_{N-1},s^z_N}$ using the following identity
\begin{align}\nonumber
\ket{j_1,j_2,j,m }=&\sum_{m_1,m_2}\ket{j_1,m_1 }\ket{j_2,m_2} W^{j_1j_2}_{m_1m_2,jm}\delta_{m,m_1+m_2}\nonumber\\
=&\sum_{m_2}\ket{j_1,m-m_2 }\ket{j_2,m_2} W^{j_1j_2}_{(m-m_2)m_2,jm}.
\end{align}
Here, $W^{j_1j_2}_{m_1m_2,jm}$ are the Clebsch-Gordan coefficients of total angular momentum eigenstates $\ket{j_1,j_2,j,m }$ in the uncoupled tensor product basis.

Therefore, we have
\begin{widetext}
\begin{align}\label{eq:derivation}
&\ket{S_1,S_{2},\cdots,S_{N-1},S_N,S^z_N}\nonumber\\
=&\sum_{s^z_N}W^{S_{N-1}\frac{1}{2}}_{(S^z_N-s^z_N)s^z_N,S_NS^z_N}\ket{S_{N-1},S^z_N-s^z_N}\ket{\frac{1}{2},s^z_N } \nonumber\\
=&\sum_{s^z_{N-1},s^z_N}W^{S_{N-2}\frac{1}{2}}_{(S^z_N-s^z_N-s^z_{N-1})s^z_{N-1},S_{N-1}(S^z_N-s^z_N)}W^{S_{N-1}\frac{1}{2}}_{(S^z_N-s^z_N)s^z_N,S_NS^z_N}\ket{S_{N-2},S^z_N-s^z_N-s^z_{N-1}}\ket{\frac{1}{2},s^z_{N-1}} \ket{\frac{1}{2},s^z_N} \nonumber\\
&\vdots\nonumber\\
=&\sum_{s^z_2,\cdots,s^z_{N-1},s^z_N}\prod_{n=2}^NW^{S_{n-1}\frac{1}{2}}_{(S^z_N-\sum^N_{k=n}s^z_{k})s^z_n,S_n(S^z_N-\sum^N_{k=n}s^z_{k}+s^z_n)}\ket{s^z_1,s^z_{2},\cdots,s^z_{N-1},s^z_N}\nonumber\\
=&\sum_{s^z_2,\cdots,s^z_{N-1},s^z_N}\prod_{n=2}^NW^{S_{n-1}\frac{1}{2}}_{\big(\sum^{n-1}_{k=1}s^z_{k}\big)s^z_n,S_n\big(\sum^n_{k=1}s^z_{k}\big)}\ket{s^z_1,s^z_{2},\cdots,s^z_{N-1},s^z_N}\nonumber\\
=&\sum_{s^z_2,\cdots,s^z_{N-1},s^z_N}\prod_{n=2}^NW^{S_{n-1}\frac{1}{2}}_{S^z_{n-1}s^z_n,S_{n}S^z_n}\ket{s^z_1,s^z_{2},\cdots,s^z_{N-1},s^z_N},
\end{align}
\end{widetext}
where we have used the conservation of total angular momentum projection in the $z$-direction, i.e., $S^z_N=\sum^N_{k=1}s^z_{k}$. Since the Clebsch-Gordan coefficients can always be chosen as real numbers, we thus can represent the uncoupled tensor state in the LMG basis
\begin{align}\label{s1sNS1SN}
&\ket{s^z_1,s^z_{2},\cdots,s^z_{N-1},s^z_N}\nonumber\\
&=\sum_{S_1,\cdots,S_N}\prod_{n=2}^NW^{S_{n-1}\frac{1}{2}}_{S^z_{n-1}s^z_n,S_{n}S^z_n}\ket{S_1,S_{2},\cdots,S_{N-1},S_N,S^z_N}.
\end{align}
This expression is the general form for the three spins Eq.~\eqref{Neel3spinsfull}. Because the flexibility of spin chain configurations $S_1,\cdots,S_{N-1}$, the subspace $\{\ket{S_N=S,S^z_N=m}\}$ is degenerate with its degeneracy $D_N(S)$ given by
\begin{align}\label{}
D_N(S)=\frac{2S+1}{N+1}{N+1 \choose \frac{N}{2}-S}.
\end{align}
Applying Eq.~(\ref{s1sNS1SN}) to the \text{N\'eel} state with an odd number of spins and collecting all the states in the degenerate subspace, we obtain a general expression for arbitrary (odd) system sizes
\begin{align}\label{NeelN}
\ket{\text{N\'eel}}_N&\equiv\underbrace{\ket{\uparrow\downarrow\cdots\uparrow}}_N\nonumber\\
&=\sum_S\sqrt{\frac{D_N(S)}{\sum_{S'} D_N(S')}}\ket{S_N=S,\frac{1}{2}}.
\end{align}
The above expression can be proved as follows. We first we assume Eq.~(\ref{NeelN}) is valid for \text{N\'eel} state of $N$ spins, which is true for $N=3$ from Eq.~(\ref{3spins}). By representing the product state $\ket{\downarrow\uparrow}$ in the basis of singlet and triplet states, we can obtain the \text{N\'eel} state of $N+2$ spins via the following construction
\begin{widetext}
\begin{align}\label{NeelN2}
\ket{\text{N\'eel}}_{N+2}
&=\ket{\text{N\'eel}}_N\otimes\ket{\downarrow\uparrow}\nonumber\\
&\propto\sum_S\sqrt{D_N(S)}\ket{S_N=S,\frac{1}{2}}\frac{1}{\sqrt{2}}\big(\ket{1,0}-\ket{0,0}\big)\nonumber\\
&=\sum_S\sqrt{\frac{D_N(S)}{2}}\Big(W^{S,1}_{\frac{1}{2},0;S-1,\frac{1}{2}}\ket{S-1,\frac{1}{2}}+W^{S,1}_{\frac{1}{2},0;S+1,\frac{1}{2}}\ket{S+1,\frac{1}{2}}+W^{S,1}_{\frac{1}{2},0;S,\frac{1}{2}}\ket{S,\frac{1}{2}}-W^{S,0}_{\frac{1}{2},0;S,\frac{1}{2}}\ket{S,\frac{1}{2}}\Big)\nonumber\\
&=\sum_S\sqrt{\frac{1}{2}\big[D_N(S+1)(W^{S+1,1}_{\frac{1}{2},0;S,\frac{1}{2}})^2+D_N(S-1)(W^{S-1,1}_{\frac{1}{2},0;S,\frac{1}{2}})^2+D_N(S)((W^{S,0}_{\frac{1}{2},0;S,\frac{1}{2}})^2+(W^{S,1}_{\frac{1}{2},0;S,\frac{1}{2}})^2)\big]}\ket{S,\frac{1}{2}}\nonumber\\
&\propto\sum_S\sqrt{D_{N+2}(S)}\ket{S_{N+2}=S,\frac{1}{2}}.
\end{align}
\end{widetext}
In the last two steps, we have regrouped the terms with the same total spin $S_{N+2}=S$ and made use of the identities
\begin{subequations}
\begin{align}
W^{S+1,1}_{\frac{1}{2},0;S,\frac{1}{2}}&=-\frac{1}{2}\sqrt{\frac{2S+1}{S+1}},\\
W^{S,1}_{\frac{1}{2},0;S,\frac{1}{2}}&=\frac{1}{2}\sqrt{\frac{1}{S(S+1)}},\\
W^{S-1,1}_{\frac{1}{2},0;S,\frac{1}{2}}&=\frac{1}{2}\sqrt{\frac{2S+1}{S}},\\
W^{S,0}_{\frac{1}{2},0;S,\frac{1}{2}}&=1.
\end{align}
\end{subequations}
By normalising Eq.~(\ref{NeelN2}), the identity Eq.~(\ref{NeelN}) is thus proved. For simplicity, we drop the footnote of $D_N(S)$ and obtain Eq.~(\ref{Neel}) in the main text. The present approach allows us to evaluate any expectation value of operators that can be expressed in terms of $\hat{S}_\alpha$, which in particular includes the Loschmidt echo and its quench dynamics. However, a staggered magnetization is not covered by this class, which is easy to understand as staggering is not uniquely defined in a zero-dimensional system.

\bibliography{afm}
\end{document}